\documentclass[useAMS,usenatbib]{mn2e}
\usepackage{graphicx}
\usepackage[fleqn]{amsmath}
\usepackage{amssymb}
\usepackage{multirow}

\def\nm{\,nm}

\title[Transmission spectroscopy of WASP-29b]{A Gemini ground-based transmission spectrum of WASP-29b: a featureless spectrum from 515 to 720\,nm}
\author[N. P. Gibson et al.]{
N. P. Gibson$^{1,2}$\thanks{E-mail: ngibson@eso.org},
S. Aigrain$^{1}$,
J. K. Barstow$^{1}$,
T. M. Evans$^{1}$,
L. N. Fletcher$^{3}$, and \newauthor
P. G. J. Irwin$^{3}$.
\smallskip
\\
$^{1}$Department of Physics, University of Oxford, Denys Wilkinson Building, Keble Road, Oxford OX1 3RH, UK\\
$^{2}$European Southern Observatory, Karl-Schwarzschild-Str. 2, 85748 Garching bei M\"unchen, Germany\\
$^{3}$Atmospheric, Oceanic and Planetary Physics, University of Oxford, Clarendon Laboratory, Oxford, OX1 3PU, UK\\
}

\begin{document}

\date{Accepted 2012 October 29}

\pagerange{\pageref{firstpage}--\pageref{lastpage}} \pubyear{2002}

\maketitle

\label{firstpage}

\begin{abstract}

We report Gemini-South GMOS observations of the exoplanet system WASP-29 during primary transit as a test case for differential spectrophotometry. We use the multi-object spectrograph to observe the target star and a comparison star simultaneously to produce multiple light curves at varying wavelengths. The `white' light curve and fifteen `spectral' light curves are analysed to refine the system parameters and produce a transmission spectrum from $\sim$515 to 720\,nm. All light curves exhibit time-correlated noise, which we model using a variety of techniques. These include a simple noise rescaling, a Gaussian process model, and a wavelet based method. These methods all produce consistent results, although with different uncertainties. The precision of the transmission spectrum is improved by subtracting a common signal from all the spectral light curves, reaching a typical precision of $\sim$1$\times10^{-4}$ in transit depth. The transmission spectrum is free of spectral features, and given the non-detection of a pressure broadened Na feature,
we can rule out the presence of a Na rich atmosphere free of clouds or hazes, although we cannot rule out a narrow Na core. This indicates that Na is not present in the atmosphere, and/or that clouds/hazes play a significant role in the atmosphere and mask the broad wings of the Na feature, although the former is a more likely explanation given WASP-29b's equilibrium temperature of $\sim$970\,K, at which Na can form various compounds. We also briefly discuss the use of Gaussian process and wavelet methods to account for time correlated noise in transit light curves.

\end{abstract}


\section{Introduction}

The study of transiting exoplanets is rapidly advancing our understanding of planets beyond our own Solar system. Planets' bulk densities are obtained via measurement of the radius and mass using transits and radial velocity measurements; this is the first step in understanding their structure and composition. However, to understand planetary systems more fully and explore their diversity we need spectroscopic measurements of their atmospheres; luckily, transiting planets allow such measurements without requiring the star and planet to be spatially resolved.

Transmission spectroscopy is a measurement of the effective size of the planet as a function of wavelength during primary transit. Due to wavelength dependent opacities in the atmosphere, the planet appears larger at wavelengths where the atmosphere absorbs or scatters light. We can therefore probe for the presence of atomic and molecular species, as well as clouds or hazes \citep{Seager_2000,Brown_2001}. Until recently, transmission spectroscopy has only been feasible using space-based telescopes, which have been tremendously successful, particularly at optical wavelengths \citep[e.g.][]{Charbonneau_2002,Pont_2008,Sing_2008,Sing_2011,Huitson_2012}. Measurements in the NIR have proved more controversial, with different groups reporting conflicting conclusions from the same dataset \citep[e.g.][]{Swain_2008, Gibson_2011}; however, advances in data analysis techniques are starting to resolve the issue \citep[e.g.][]{Gibson_2012,Waldmann_2012}, along with the availability of more stable NIR cameras such as WFC3 \citep[e.g.][]{Berta_2012, Gibson_2012b}.

Most recently, the use of a multi-object spectrograph (MOS) to perform differential spectro-photometry has started to show that ground based observations can also play an important role. This technique was pioneered by \citet{Bean_2010}, using the VLT to produce a transmission spectrum of the super-Earth GJ-1214b. Such observations have benefitted not only from advances in observing strategy, but also the availability of new bright targets with nearby comparison stars. Similar observations have since been made in the near-infrared \citep{Bean_2011} and with long-slit spectrographs \citep{Sing_2012}. Here we report Gemini observations of WASP-29 using this MOS technique.

WASP-29b \citep{Hellier_2010} is a Saturn-size planet with a mass and radius of $0.24\pm0.02\,M_J$ and $0.79\pm0.05R_J$. It orbits a K4 dwarf with a period of $\sim$3.9\,days. Given its relatively low equilibrium temperature of $\sim$980\,K, it does not have a particularly large scale height compared to the hottest and lowest-density hot Jupiters. It was in fact chosen as a test case for the Gemini GMOS instrument, to test its stability before more favourable targets are observed. WASP-29 has a nearby, bright comparison star, making it ideal for differential photometry or spectroscopy.

Despite its recent successes, MOS differential spectroscopy, like other techniques, is expected to suffer from correlated noise originating from poorly understood instrumental effects. Although the differential nature of the measurements is intended to reduce this problem, systematic effects are not yet fully understood for ground based differential photometry, and the extra complexity in spectroscopic work may introduce more systematic effects.
In this paper, we explore the use of several methods to analyse transit light curves to obtain useful measurements in the presence of significant correlated noise. These methods include the wavelet method of \citet{Carter_2009}, and the Gaussian process (GP) model of \citet{Gibson_2012} adapted to model time-correlated noise.

This paper is structured as follows: Sect.~\ref{sect:observations} presents our observations and data reduction procedure, in Sect.~\ref{sect:analysis} we present our transit analysis methods, and finally in Sects.~\ref{sect:results} and \ref{sect:discussion} we present our results and conclusions.

\section{GMOS Observations and Data Reduction}
\label{sect:observations}

A transit of WASP-29 was observed using the 8-m Gemini-South telescope with the Gemini Multi-Object Spectrograph (GMOS) on October 19 2011. Data were taken as part of program GS-2011B-Q-13 (PI. Gibson). GMOS has an imaging field-of-view of 5.5$\times$5.5 arcmin squared, and consists of three 2048$\times$4608 pixel CCDs arranged side by side with small gaps in-between. We used GMOS in multi-object mode to observe the target plus two comparison stars simultaneously and continuously for $\sim$5.1 hours, covering the 2.66 hour transit plus 1.5 hours prior to ingress and 1.0 hours after egress. Conditions were photometric for the duration of the observations.

Observations used the R400 grism + OG515 filter with a central wavelength of 725\nm. The dispersion is 0.07\nm\ per (unbinned) pixel, giving wavelength coverage from $\sim$515--940\nm. Exposure times were 30.5 seconds and 2$\times$2 binning was used. The full frame readout of the GMOS chips in 2$\times$2 binning is $\sim$55 seconds (in the recommended `slow' readout). In order to reduce the overheads we read out only three regions of interest (ROI) on the chip containing the three stellar spectra, resulting in a readout time of $\sim$22 seconds, and therefore a cadence of $\sim$53 seconds. This allowed 348 exposures over the 5.1 hours of observation. To minimise slit losses we created a mask with slits of 30 arcsec length and 10 arcsec width for the three stars, giving seeing limited (therefore variable) resolution ranging from R$\sim$430--860 at 700\nm.
Calibrations were taken before and after the science observations, and consisted of flat fields and arc lamp exposures. Further arcs were taken with a calibration mask. This was almost identical to the science mask but with narrow slits of 1 arcsec to enable more precise wavelength calibration.

The data were reduced using the standard GMOS pipeline contained in the Gemini {\sc IRAF}\footnote{{\sc IRAF} is distributed by the National Optical Astronomy Observatory, which is operated by the Association of Universities for Research in Astronomy (AURA) under cooperative agreement with the National Science Foundation}/{\sc PyRAF}\footnote{{\sc PyRAF} is a product of the Space Telescope Science Institute, which is operated by AURA for NASA} package. First the ROI images were processed to be in the standard GMOS format. Basic reductions included bias subtraction and wavelength calibration. Fringing is particularly large at the red end of the spectra\footnote{Fringing is not so problematic using the GMOS-North detectors, where future observations are planned.} ($>$750\nm). We tried correcting for this using the flat fields, however this proved particularly problematic given that our spectral flat fields were taken with slit widths much larger than the PSF of the star.
Flat fields taken with the calibration mask were corrupted. Given this, we decided not to apply flat-fielding. In differential spectrophotometry, flat fielding is not necessary providing the sensitivity ratios between the target and comparison stars in each wavelength channel remain constant. Of course, this is often not the case if there is movement of the spectra on the CCDs, seeing variations, etc. However, correcting all the data using the same flat field cannot account for such effects anyway (although it might mitigate against them), given spectral flat-fields are PSF (at least when using wide slits) and wavelength dependent. Simple flat-fielding is also unable to correct for severe fringing at the level required for transmission spectroscopy.

\begin{figure}
\includegraphics[width=85mm]{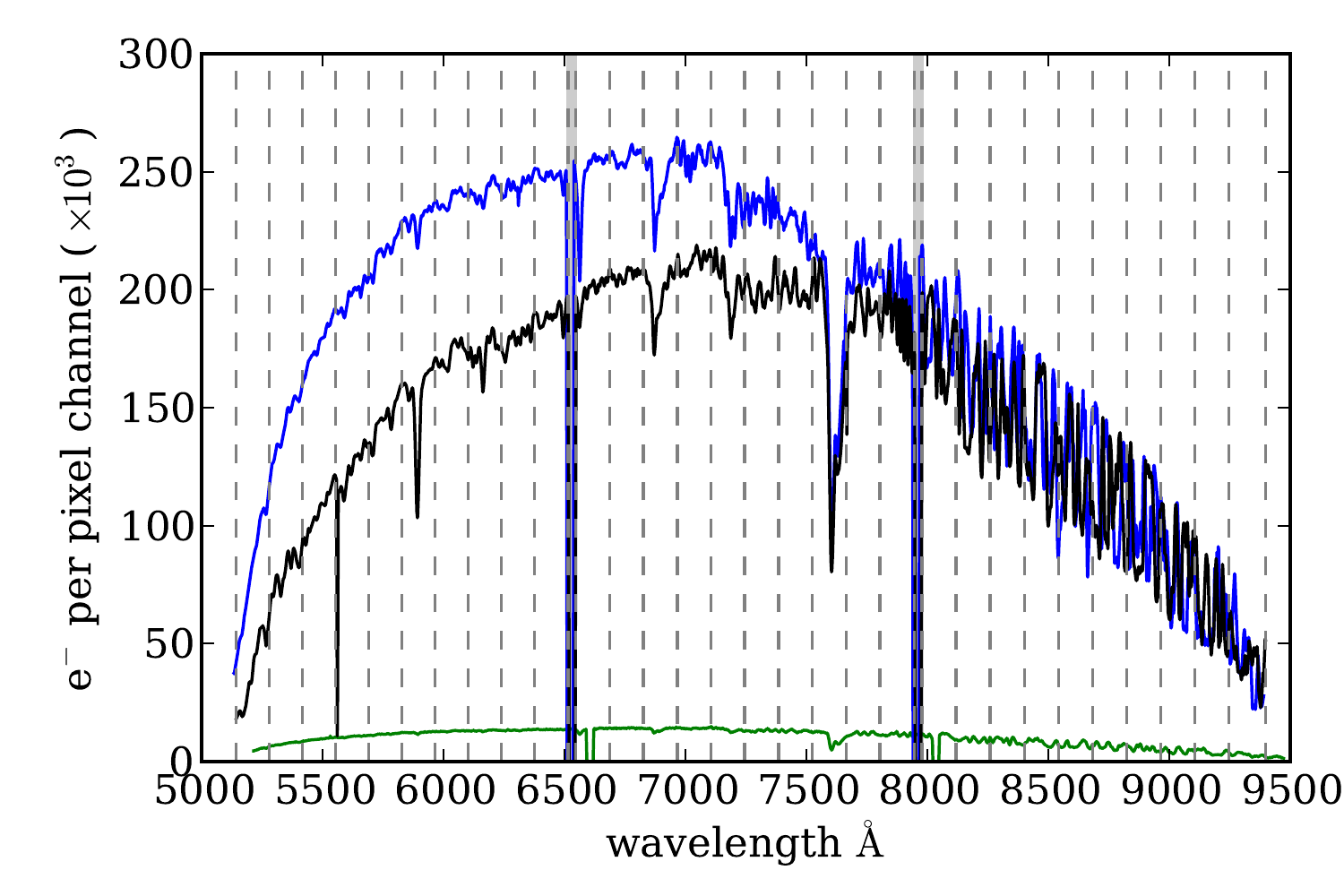}
\caption{Example spectra extracted from a single exposure. The black line is WASP-29, and the blue and green lines are the two comparison stars. The vertical dashed lines mark the extraction regions, with the shaded regions marking the gaps in the target spectrum between the detectors. Only the first 15 channels were used in the final analysis, due to significant fringing at the red end.}
\label{fig:eg_spec}
\end{figure}

Wavelength calibrated spectra for the 3 stars were extracted using the {\sc gsextract} routine, with an aperture of 8 pixels (varying the aperture by a few pixels had little effect) after sky subtraction. A few pixel columns (the spatial direction) showing significant temporal variation were masked from the extraction.
Examples of extracted spectra of WASP-29 and the two comparison stars are shown in Fig.~\ref{fig:eg_spec}, showing uncorrected fringing effects at the red end. The spectra were divided up into several spectral regions, defining independent wavelength channels. To extract the light curves for each wavelength channel, the spectra were summed in each of the spectral regions to produce a flux time-series of the target and comparison stars for each channel. The target flux was then divided through by the sum of the comparison stars' flux to produce multiple light curves at each wavelength channel, which we hereafter refer to as the `spectral' light curves. We experimented with different numbers of wavelength channels, which affect the resolution and signal to noise, and finally settled on 30 across the whole spectral range, which are marked by the vertical dashed lines in Fig.~\ref{fig:eg_spec}. We divided each of the three chips into 10 channels each of the same pixel widths, ignoring the gaps between detectors in the images.

\begin{figure}
\includegraphics[width=85mm]{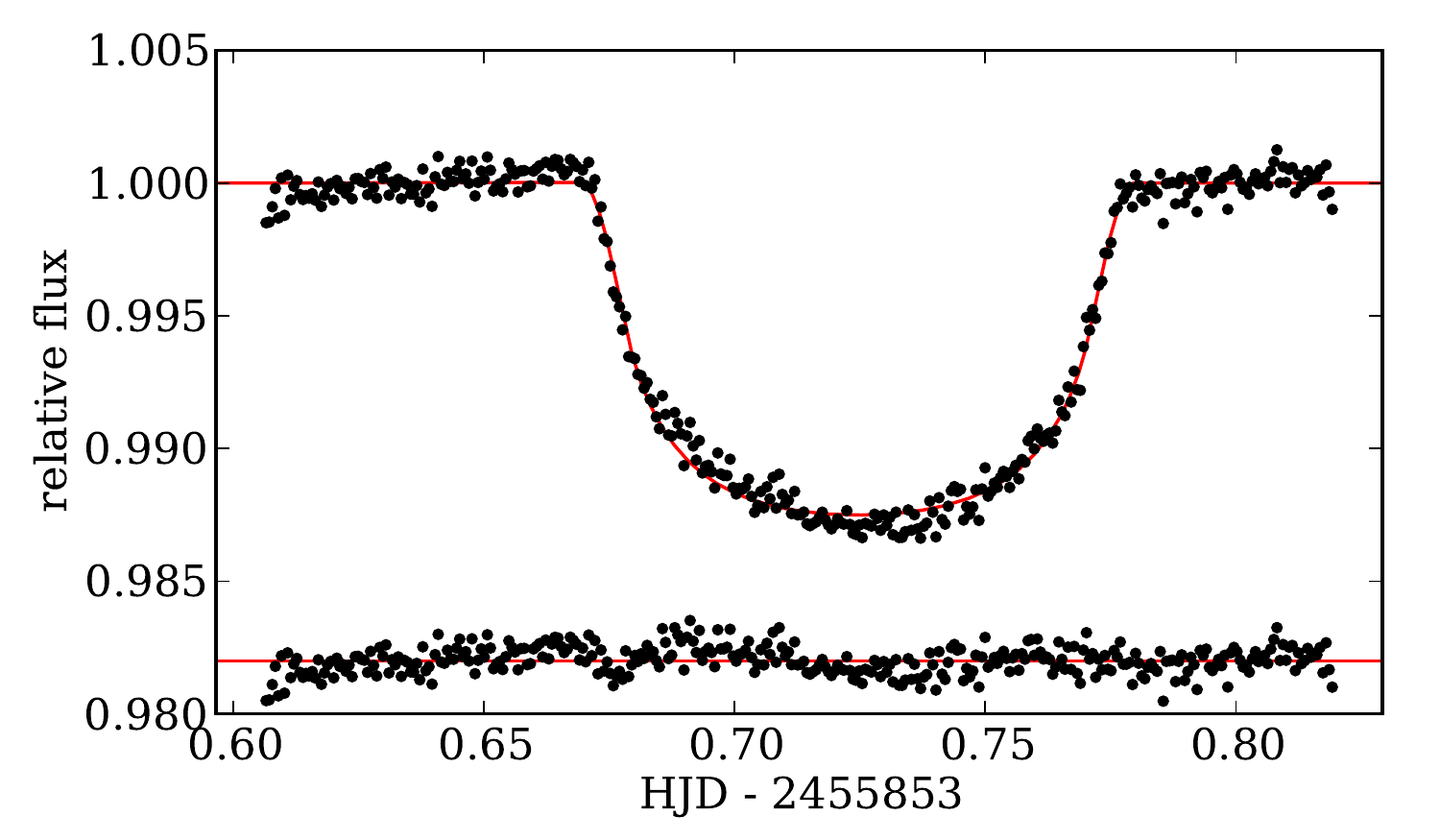}
\includegraphics[width=85mm]{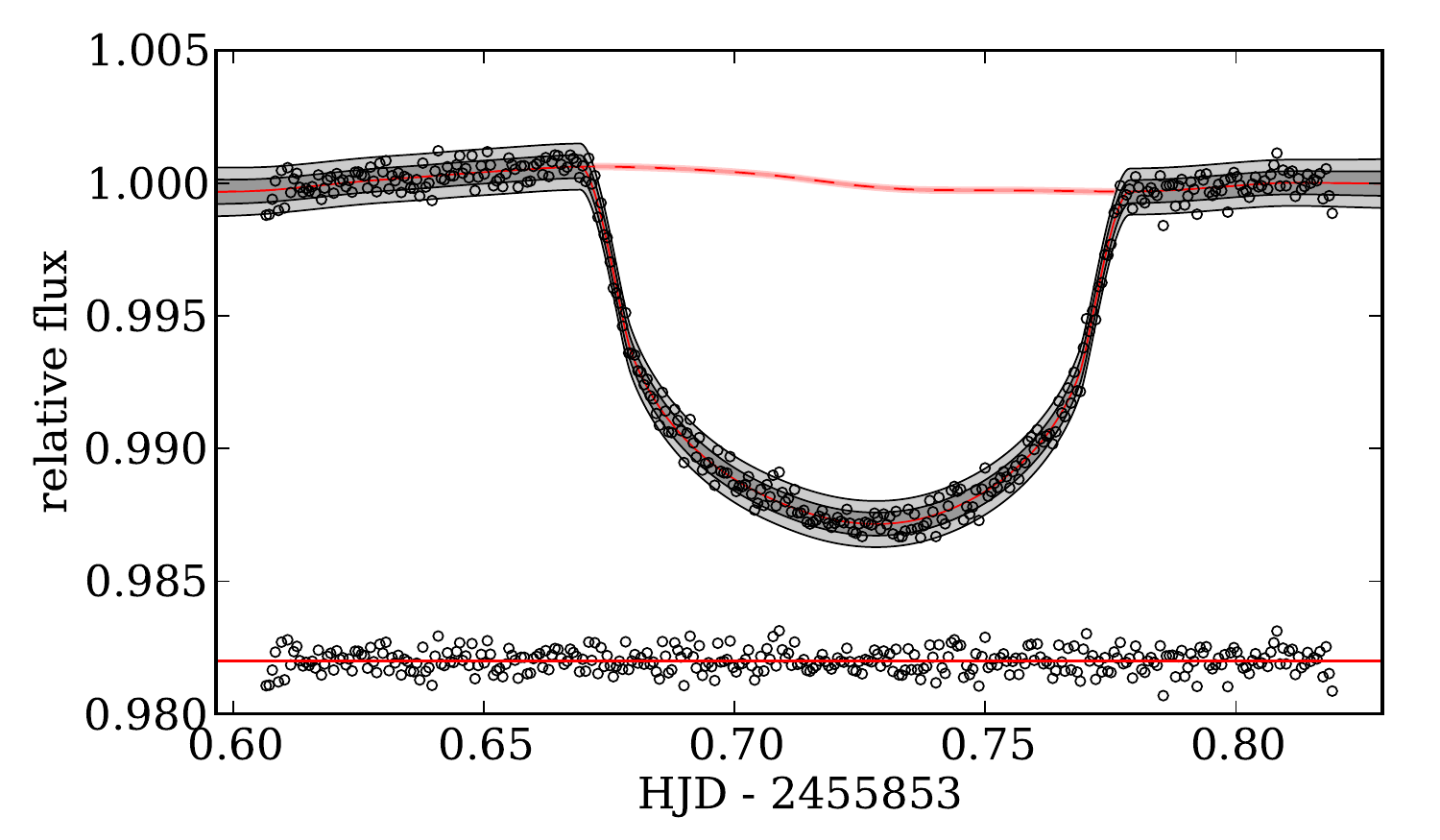}
\caption{The `white' light curve produced from the first 10 wavelength channels of the GMOS spectra. Top: best fit model using a white noise analysis and the residuals. Clearly, significant systematics are present and must be accounted for in the fitting process. Bottom: best fit using a Gaussian process model with a Mat\'ern 3/2 covariance kernel (see text). The red line represents the best fit model, with the grey regions representing the 1 and 2\,$\sigma$ limits of the instrument model plus white noise times the transit model. The dashed red line is the projection of the systematics model without the transit, along with the 1 and 2\,$\sigma$ limits (now excluding the white noise term).}
\label{fig:GP_white_lcv}
\end{figure}

\begin{figure*}
\includegraphics[width=170mm]{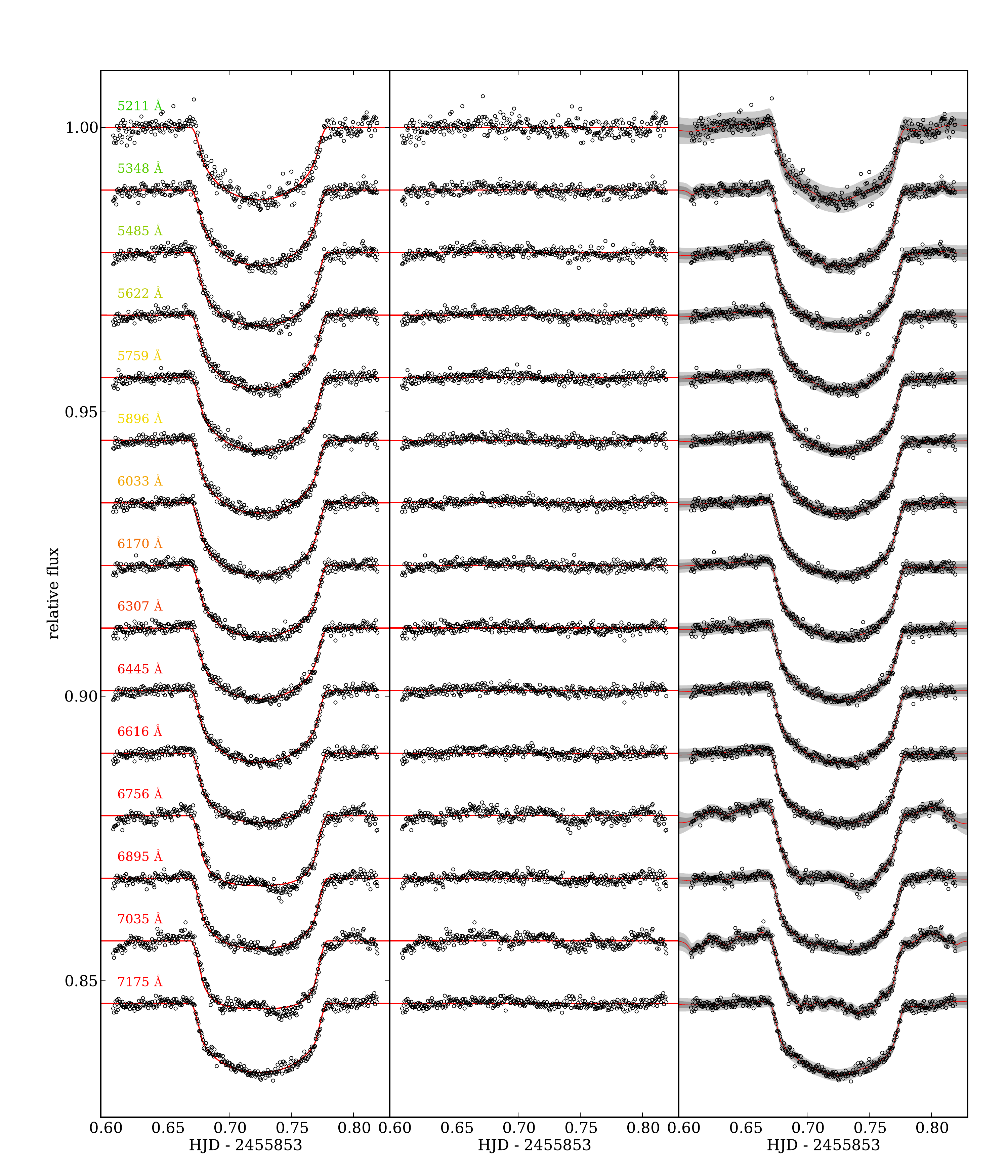}
\caption{Spectral light curves with the central wavelengths marked, with a linear trend in time removed. Left panel: light curves fitted with the simple white noise model.  Middle panel: residuals from the white noise fits. Right panel: same light curves as left panel with their best fit Gaussian process model in red. The grey shading represents the 1 and 2\,$\sigma$ limits of the GP model (including white noise).}
\label{fig:all_lightcurves}
\end{figure*}

Correlated noise is present to some degree in all the light curves, but is particularly bad at the red end due to the fringing (and possibly the strong O$_2$ telluric feature at $\sim$7590\,\AA). We therefore decided to analyse only 15 spectral light curves at the blue end and discard the remaining 15 for the remainder of this analysis. We also experimented with the two comparison stars. Given that the second comparison star is significantly fainter, it was excluded from the analysis and we only used the brighter one. Finally, we also produced a `white' light curve, by summing up the flux over the first 10 wavelength channels (the other 5 contained larger systematics), prior to dividing through the target flux by the single comparison star's flux. The white light curve is shown in Fig.~\ref{fig:GP_white_lcv}, and the spectral light curves are shown in Fig.~\ref{fig:all_lightcurves}.

We calculated the theoretical photon noise for the white light curve and each spectral bin, taking into account the read noise and sky contribution (although negligible for high signal-to-noise data). Typical integrated electron counts per exposure for each spectral bin range from $\sim$3.7$\times10^6$ to 2.1$\times10^7$ for the target star, and $\sim$7.9$\times10^6$ to 2.6$\times10^7$ for the comparison star, and vary slightly throughout the night with airmass. For the white light curve the typical integrated counts per exposure were $\sim$1.3$\times10^8$ and 2.0$\times10^8$ electrons for the target and comparison star, respectively. This results in (time-averaged) theoretical precision on the relative flux per exposure ranging from 2.9$\times10^{-4}$ to 6.2$\times10^{-4}$ for the spectral light curves, and $\sim$1.1$\times10^{-4}$ for the white light curve.

We also extracted auxiliary data from our target and comparison star's spectra to investigate the cause of the systematics present in the light curves. The relative shift in the dispersion axis was measured by cross correlating each star's spectra with the first in the time-series. The relative shift in the cross-dispersion axis and the width of the spectral trace were found by fitting a Gaussian function to each column of the spectrum, and finding the average values per exposure. Over the course of the observations, the shift in the dispersion and cross-dispersion axes were $\sim$1.0 and 0.5 (binned) pixels respectively, and were the same for both stars (within the measurement error of $\sim0.1$ pixels), showing that no significant rotation of the field occurred. The FWHM in the cross dispersion direction ranged from 6--12 (binned) pixels, resulting in varying spectral resolution. We found no obvious correlations between these measurements and the systematics in the light curves, although the seeing variations are likely the most significant contributor given the relatively large changes.

\section{Light Curve Analysis}
\label{sect:analysis}

The white light curve and the 15 spectral light curves were modelled in a variety of ways to account for the instrumental systematics. We first analysed them using a simple white noise model to inspect the residuals and establish the nature of the systematics. In all cases the transit model was constructed using the analytic equations of \citet{mandel_agol_2002}, and is similar to that described in \citet{Gibson_2008}. We assumed a circular orbit, and fixed the period ($P$) to 3.922727 days as given in \citet{Hellier_2010}. The remaining parameters of the transit model were the central transit time ($T_\text{C}$), the system scale ($a/R_\star$), the planet-to-star radius ratio ($\rho=R_\text{p}/R_\star$), the impact parameter ($b$), the two quadratic limb darkening parameters ($c_1$,$c_2$), and two parameters of a linear baseline model of time ($f_\text{oot}$, $T_\text{grad}$). We hereafter denote the transit model as $T(\bmath t,\bphi)$, where $\bmath t$ is a vector of time, $\bphi$ is the vector of transit parameters, and $\bmath f$ as the vector of flux measurements.

Unless otherwise stated, all of these parameters except $P$ were fitted to the white light curve, and $\rho$, $c_1$, $c_2$, $f_\text{oot}$ and $T_\text{grad}$ were fitted to the spectral light curves with the remaining parameters fixed to the {\it final} white light curve values (see Sect.~\ref{sect:results_wlc}). This is because we are interested in finding the relative values for $\rho$ as a function of wavelength to produce our transmission spectrum, and we can condition on the values of $T_\text{C}$, $a/R_\star$ and $b$, that change the inferred values for $\rho$ in the same way.

Given the lack of correlations found between the systematics and the auxiliary information extracted (see Sect. \ref{sect:observations}), we model the systematics as time-correlated noise. The only differences in the following analyses are the noise models used to account for instrumental systematics. These are described in turn.

\subsection{White noise analysis}
\label{sect:whiteanalysis}

We first analysed all of the light curves using a simple white noise model, with likelihood given by
\begin{equation}
\label{eq:white_noise_likelihood}
p(\bmath f| \bmath t,\bphi) = \mathcal{N} \left (T(\bmath t,\bphi) , \sigma_\text{w}^2\mathbfss{I} \right),
\end{equation}
where $\mathcal{N}(\bmu,\mathbf\Sigma)$ is the multivariate normal distribution with mean $\bmu$ and covariance matrix $\mathbf\Sigma$, $\sigma_\text{w}$ is the uncertainty of each data point and \mathbfss{I} is the identity matrix. In other words we have a diagonal covariance matrix with all diagonal terms equal to $\sigma_\text{w}^2$, representing a standard i.i.d. noise model as is commonly used to model transits.

We then multiply the likelihood by the prior on the transit and noise parameters, $p(\bphi,\sigma_\text{w}^2)$, to produce the joint posterior probability distribution, and use a Monte-Carlo Markov-Chain (MCMC) to explore the posterior distribution and produce marginal probabilities for each of the model parameters. In practice we do not explicitly state priors for most parameters, implying uniform, improper priors. The exceptions are for the limb darkening parameters and impact parameter\footnote{strictly speaking we should apply the same prior to $\rho$ and $a/R_\star$, but this would have no effect on the inference.}, where we restrict the parameter to be positive using a step function of the form;
\[
p(x) = \left\{
\begin{array}{l}
0,~\mathrm{if}~x < 0 \\
1,~\mathrm{if}~x \geq 0 \\
\end{array}
\right.,
\]
specifying another improper prior. We also restrict the sum of the two limb darkening parameters $c_1+c_2 \leq 1$ in a similar way, to ensure that the brightness of the stellar surface is positive. Four MCMC chains were run for each light curve, of length 100\,000. We excluded the first 10\% of each chain, and verified convergence by checking the Gelman \& Rubin (GR) statistic \citep{GelmanRubin_1992}. The light curves along with their best fit models are shown in Figs.~\ref{fig:GP_white_lcv} and \ref{fig:all_lightcurves} for the white and spectral light curves, respectively. 

\begin{figure}
\includegraphics[width=85mm]{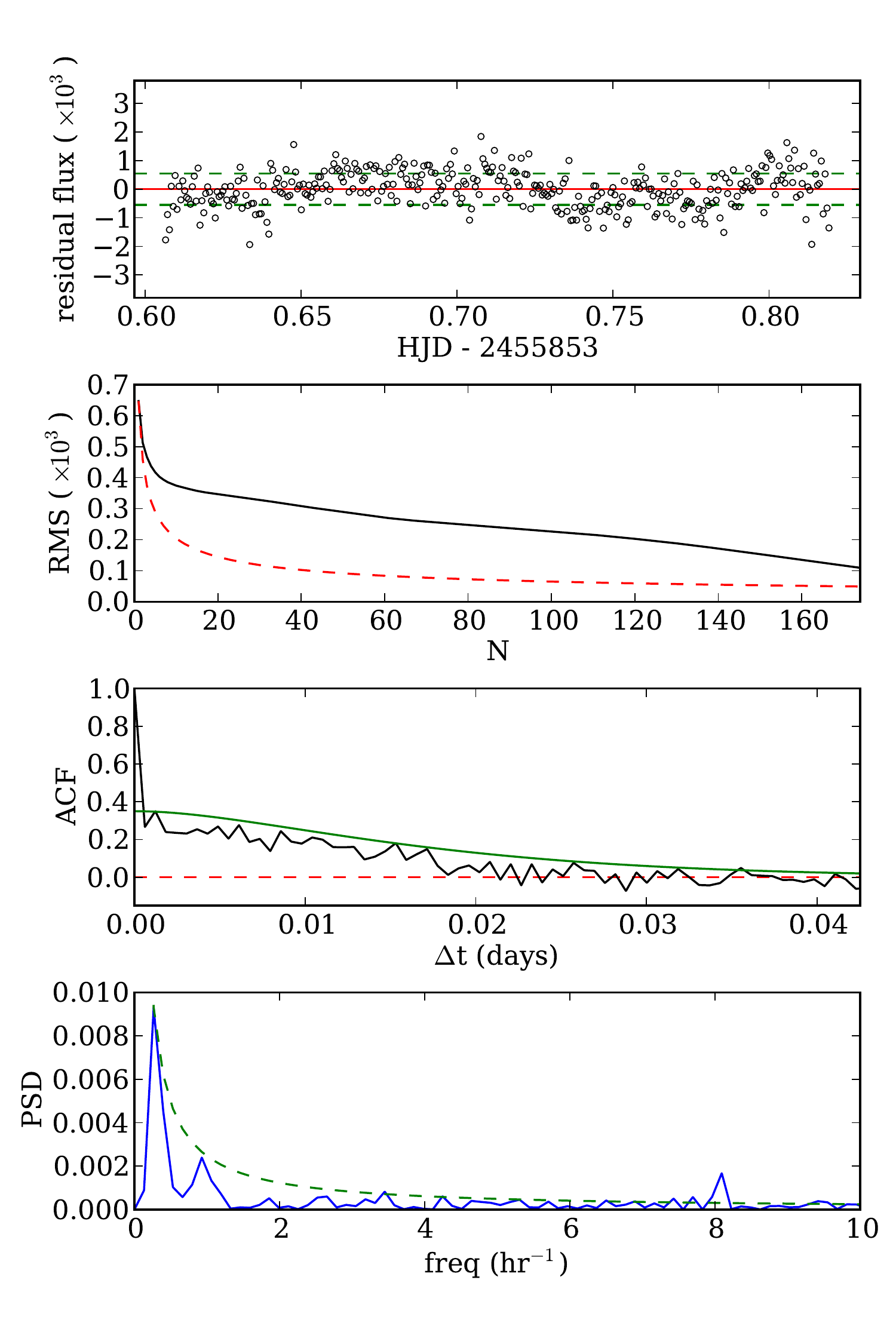}
\caption{Residuals analysis of one of the spectral light curves (6895\,\AA).
Top panel: Residuals from the best fit white noise model. The dashed lines represent the white noise RMS.
Second panel: Plot of the RMS as a function of bin size $N$. The dashed line represents the theoretical curve expected for white noise, and the ratio of the two curves gives the factor $\beta$, used to re-scale the uncertainties for the simple rescaling systematics model.
Third panel: Autocorrelation function of the residuals. The solid green line marks the best fit Mat\'ern 3/2 kernel from the GP model, and the dashed red line zero correlation.
Bottom panel: Power spectral density of the residuals. The dashed line marks a $1/f$ envelope.
}
\label{fig:residuals_analysis}
\end{figure}

Clearly, the white noise model is incapable of accounting for the correlated noise in the light curves, as seen in the residuals. We therefore analyse the residuals of each of the best fit models in an effort to understand the form of the systematics. We first use the time-averaging method to obtain a simple estimate of the red noise \citep{Pont_2006}, following the procedure of \citet{Winn_2008}, where the residuals are averaged into bins of width $N$, and the RMS is calculated as a function of N. The noise should drop by $1/\sqrt{N}$ if it is uncorrelated in time\footnote{actually, $\frac{1}{\sqrt{N}}\sqrt{\frac{M}{M+1}}$, where $M$ is the number of bins}. See \citet{Gibson_2009} for a more detailed description of this procedure. Fig.~\ref{fig:residuals_analysis} shows an example of this for one of the light curves, clearly showing that there is time-correlated noise in the light curves. As a first attempt to account for correlated noise, we calculate the factor $\beta$, which is the ratio of the RMS vs $N$ plot to the theoretical noise in the white case. We chose the maximum value for this, and then scale the noise parameter, $\sigma_\text{w}$, by this value, fix it, and re-fit the light curves using artificially inflated error bars to account for the systematics using the same MCMC procedure as before. $\beta$ ranged from $\sim$2.4 to 3.8. We will hereafter refer to this as the `white noise plus $\beta$' model.

This method is a useful way to estimate the additional uncertainties expected in the presence of systematic noise. However, it does not allow the form of the correlations to be modelled and therefore cannot produce more accurate parameter estimates (this is discussed in \citealt{Carter_2009}), and therefore we consider more sophisticated models in the following sections.

\subsection{Gaussian process analysis}
\label{sect:GPanalysis}

We use a Gaussian process to model the time-correlated noise, similar to that described in \citet{Gibson_2012} and \citet{Gibson_2012b}, allowing us to model instrumental systematics as a stochastic rather than a deterministic process. This avoids the need to specify a parametric form for the systematics, which is often impossible to do, whilst also allowing for a much more flexible model. Furthermore, GPs are intrinsically Bayesian, thus avoiding the possibility of over-fitting systematics. The combination of a very flexible model and principled Bayesian inference effectively allows one to marginalise out any ignorance about the form of the systematics model and account for it whilst inferring transit parameters. This is extremely challenging to achieve using parametric models as it requires calculation of the Bayesian evidence (and therefore proper, usually informative, priors on the model parameters), and perhaps even marginalisation over many possible instrument models.

Here we briefly describe the GP model, and refer to \citet{Gibson_2012} and references therein for further details. A GP is a collection of random variables, any finite subset of which have a joint Gaussian distribution. Therefore we can write our GP as 
\begin{equation}
\label{eq:gp}
p(\bmath f| \bmath t,\bphi,\btheta) = \mathcal{N} \left (T(\bmath t,\bphi) , \mathbf{\Sigma} (\bmath t,\btheta) \right).
\end{equation}
The only difference to Eq.~\ref{eq:white_noise_likelihood} is that we now consider the off diagonal elements of the covariance matrix. As well as specifying a {\it mean function}, in this case the transit function, with a GP model we must also specify a {\it kernel function} which populates the elements of the covariance matrix and has hyperparameters\footnote{A Gaussian process, $\mathcal{GP}(\bmu,\mathbf\Sigma)$, is fully specified by its mean $\bmu$ and covariance $\mathbf\Sigma$. Parameters of both the mean function and kernel are known as hyperparameters.} $\btheta$, written as:
\[
\mathbf\Sigma_{nm} = k({t}_n,{t}_m | \btheta).
\]
We will discuss the choice of the kernel function later.

The above GP uses the kernel to model the {\it residuals} from the light curve model. Alternatively, we can model the light curve as the transit light curve {\it times} a GP, if we wish our systematics model to be multiplicative rather than additive, i.e.:
\[
\bmath f = T(\bmath t,\bphi, \btheta) \times \mathcal{GP}(\mathbf 1, \mathbf{\Sigma} (\bmath t,\btheta)).
\]
In this case the joint probability distribution can be written for $\bmath f / T(\bmath t,\bphi)$ as
\[
p(\bmath f / T(\bmath t,\bphi, \btheta) | \bmath t,\bphi) = \mathcal{N} \left (\mathbf 1 , \mathbf{\Sigma} (\bmath t,\btheta) \right).
\]
In practice it makes little difference which model we choose, given that transit light curves are shallow, and we could also combine multiplicative and additive GPs. For the remainder of this paper we use the latter, multiplicative model.

We tested several different types of kernels to model the time-correlated noise, including the squared exponential, Mat\'ern and rational quadratic (see \citealt{Rasmussen_Williams} for a detailed discussion of kernels, which is beyond the scope of this paper). In a fully Bayesian analysis, we could calculate the Bayesian evidence for each kernel, and use it to choose the best kernel, or alternatively even marginalise over them. However, not only is this computationally prohibitive, but it would also require us to specify proper (therefore informative) priors on the hyperparameters, which would make the evidence somewhat subjective. We therefore selected a kernel based on analysis of the white noise model residuals, and by running tests on simulated light curves.

In the end we decided to use a Mat\'ern 3/2 kernel function, given by:
\begin{equation}
\label{eq:kernel}
k({t}_n, {t}_m | \btheta) = \xi^2 \left( 1+{\sqrt{3}\eta\Delta t} \right) \exp \left( -{\sqrt{3}\eta\Delta t}\right) + \delta_{nm}\sigma_\text{w}^2,
\end{equation}
where $\xi$ is a hyperparameter that specifies the maximum covariance, $\Delta t = |t_n-t_m|$ is the time difference, $\eta$ is the inverse characteristic length scale, and $\delta$ is the Kronecker delta. This kernel can be seen as a rougher version of the commonly used squared exponential kernel (i.e. that used in \citealt{Gibson_2012,Gibson_2012b}). One major motivation for this kernel is that is gives the best match to the auto-correlation function (ACF) for most of the residuals from the white noise model. This provides an estimate of how the data points are correlated with one another as a function of time lag, and an example is shown in Fig.~\ref{fig:residuals_analysis}. The green line in the ACF function marks the best fit covariance kernel (although in practice we marginalise over the kernel parameters). We also ran a series of tests on simulated light curves, described in the Appendix, which further validate the use of the Mat\'ern kernel. We stress that selecting a kernel, whilst in some ways analogous to selecting a parametric model, allows a much more flexible model than any parametric form and is also intrinsically Bayesian. In addition, we ran much of the same analysis using the squared exponential kernel. This gave similar results, and we therefore conclude that the choice of kernel is not critical for this particular dataset.

Analogously to the white noise case, we can specify priors for all the hyperparameters of the model, and multiply by the marginal likelihood\footnote{called the marginal likelihood because in a GP we have marginalised over all the possible functions for each set of hyperparameters} to produce the posterior joint probability distribution. This can then be optimised with respect to the hyperparameters using a Nelder-Mead simplex algorithm, or alternatively the marginal parameter distributions for each hyperparameter can be obtained by exploring the posterior distribution with an MCMC in just the same way as for the white noise model. The same priors were applied to the limb darkening parameters and $b$, and we also specified hyperpriors for the hyperparameters $\xi$ and $\eta$. These took the form of Gamma distributions with shape parameter unity, given by
\[
p(x) = \left\{
\begin{array}{ll}
0, & ~\mathrm{if}~x < 0 \\
\dfrac{1}{l} \exp\left(-x/l\right), & ~\mathrm{if}~x \geq 0 \\
\end{array}
\right.,
\]
where $l$ is the length scale of the hyperprior. We set the length scales for $\xi$ and $\eta$ as $10^{-3}$ and $200$, respectively. These were not specified to influence the results of the inference, but rather to ease convergence of the MCMC chains (when both parameters are small they are unconstrained by the likelihood). Indeed, we checked that the length scales of the hyperpriors did not affect the inferred transit parameters.

Whilst GPs are rather simple in theory, each evaluation of the marginal likelihood requires inversion of the covariance matrix, which makes a full marginalisation tedious as it requires $\mathcal{O}(n^3)$ computations, and limits full GP analyses to relatively small datasets. Therefore in addition to the full GP marginalisation, we also use a technique called Maximum Likelihood type-II (ML-II), where the hyperparameters of the covariance kernel are fixed to their maximum-posterior values, and we marginalise over the remaining transit parameters. Once the covariance hyperparameters are fixed this negates the need to invert the covariance matrix. This is a valid approximation when the posterior is sharply peaked with respect to the covariance hyperparameters, and is particularly useful when running many tests on the data, although for our final results we always use fully marginalised GPs (i.e. we marginalised over the covariance hyperparameters as well).

\begin{figure*}
\includegraphics[width=170mm]{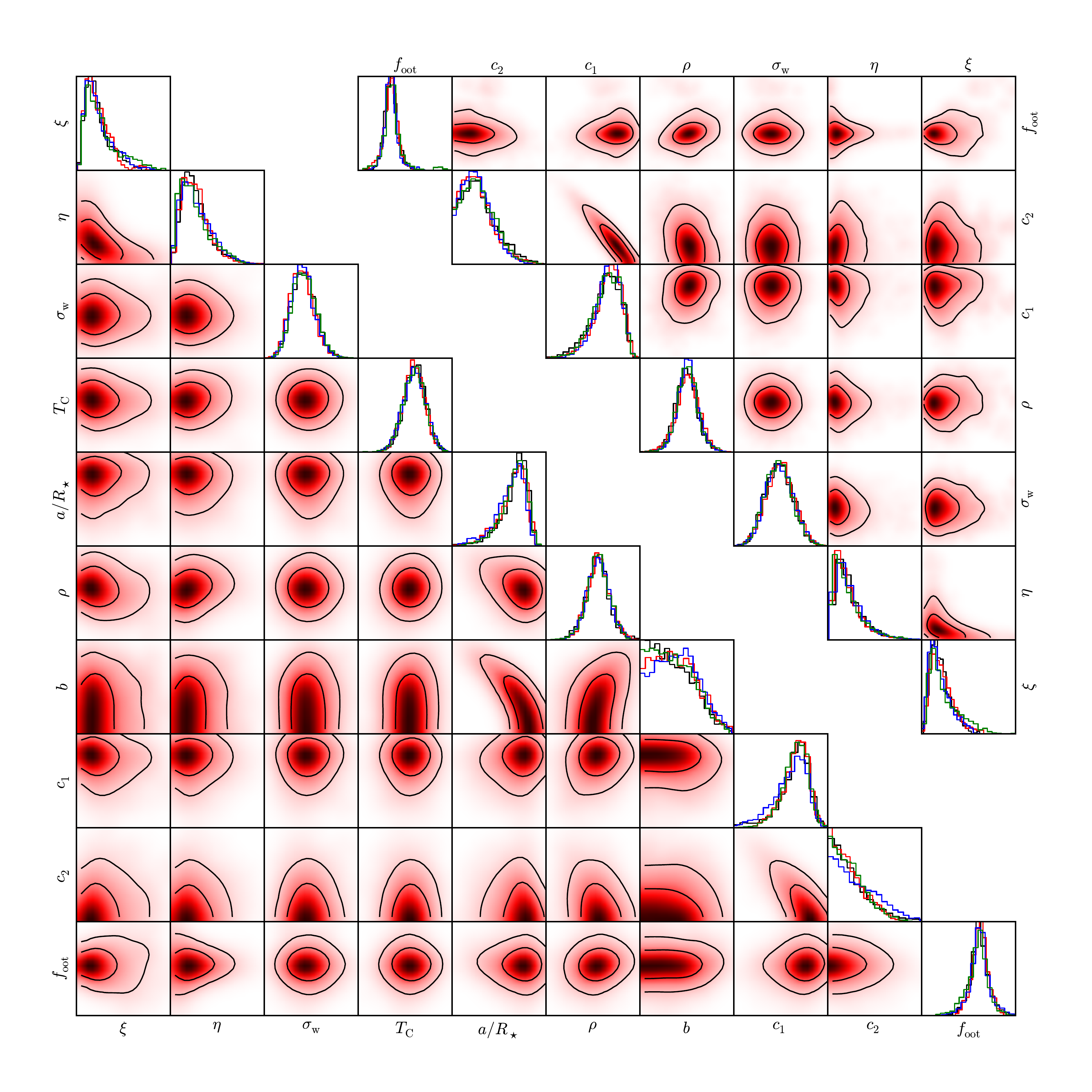}
\caption{1D and 2D marginal distributions from the posterior probability distribution for the Gaussian process noise model. The lower left shows the probability distributions for the white light curve, and the upper right for one of the spectral light curves. The black lines in the 2D distributions mark the 1 and 2\,$\sigma$ limits, respectively. Distributions for the four separate MCMC chains are shown in the 1D histograms.}
\label{fig:wlc_correlations}
\end{figure*}

We ran four chains for all light curves of length 100\,000 and 50\,000 for the ML-II and full marginalisations, respectively. Each chain of length 50\,000 for the full marginalisation took about 17.5 minutes to compute using a single core on a standard desktop, compared to about 2 minutes and 1 minute for 100\,000 length chains with the ML-II method and white noise model, repectively. The run times do not scale as badly as one might expect with $\mathcal{O}(n^3)$ complexity, because the computation of the light curve model rather than the likelihood dominates for the simpler noise models. We tested for convergence in the same way as before. Due to degeneracy in the linear baseline and the GP model, we decided to fix the $T_\text{grad}$ parameter for the full GP marginalisation. This requires shorter chains to reach convergence, and doesn't effect the results as $\rho$ did not significantly correlate with $T_\text{grad}$. We verified this with longer chains for a subset of the light curves. In a few cases, $\xi$ and/or $\eta$ did not fully converge (the GR statistic was a few percent from unity), but only when one or both parameters were consistent with zero; however, in all cases $\rho$ converged and therefore the transmission spectrum is not affected. The best fit GP models to the light curves are shown in Figs.~\ref{fig:GP_white_lcv} and \ref{fig:all_lightcurves}. To illustrate correlations in the parameters and convergence of the final marginal distributions, 1D and 2D marginal distributions are shown in Fig.~\ref{fig:wlc_correlations} for the white light curve and one spectral light curve.

\subsection{Wavelet analysis}
\label{sect:waveletanalysis}

As we have used a GP to fit for time-correlated noise only, it is worth exploring other methods in the literature to account for time-correlated noise. One such method is the wavelet method introduced by \citet{Carter_2009}. It is valid when the power spectral density (PSD) of the noise takes the form:
\[
\mathrm{PSD} \propto \frac{1}{f^\gamma},
\]
where $f$ is the frequency and $\gamma$ is the spectral index. The method is based on taking a wavelet transform of the residuals from the best fit model, and the likelihood is computed from the wavelet coefficients. The central idea of this method is that it diagonalises the covariance matrix of Eq.~\ref{eq:gp} when specified in the wavelet domain. Like GPs it models the systematics as a stochastic process; however, it has the significant advantage that a burdensome matrix inversion is not required for each evaluation of the likelihood and is therefore much faster. The extra computation required as compared to the white noise analysis is a fast wavelet transform, which is of order $\mathcal{O}(n)$, and therefore does not significantly extend the computation time.

In order to justify the wavelet method we analysed the PSD of all the residuals from the white noise fit as recommended by \citet{Carter_2009}. An example of this is shown in Fig.~\ref{fig:residuals_analysis}. In most cases the noise appeared consistent with $1/f^\gamma$, by which we mean that low frequency components dominate, but the shape itself is hard to determine (the PSD should appear as noise within a $1/f^\gamma$ envelope). This supports the use of the wavelet based likelihood for our light curves, or at least suggests the noise properties are nearly $1/f^\gamma$.

Our likelihood therefore took the form of Eq.~32 from \citet{Carter_2009}, and we multiplied by priors similarly to Sect.~\ref{sect:whiteanalysis} to produce a posterior distribution. In addition to the white noise parameter $\sigma_\text{w}$, the wavelet likelihood has a red noise parameter $\sigma_\text{r}$ specifying the amplitude of the red noise component, and the spectral index $\gamma$. We fixed $\gamma$ to 1 (as in \citealt{Carter_2009}), and optimised and explored the joint posterior distribution with respect to the free transit parameters, plus $\sigma_\text{w}$ and $\sigma_\text{r}$. Chains lengths and run times were approximately the same as for the white noise analysis.

\subsection{Removal of common mode systematics}
\label{sect:commonmode}

Finally, we experimented with a simple method to remove common signals observed in the spectral light curves, to see if we could increase the precision in our transmission spectrum.
We were motivated to do this as we obtain a reduced $\chi^2$ significantly lower than one for our transmission spectrum using all of the noise models (see Sect.~\ref{sect:trans_spectrum}). This is extremely unlikely to happen by chance, and is probably the result of similar systematic signals in the spectral light curves. Noise models will take all signals into account when calculating the uncertainties in $\rho$ (as they should); however, common signals will increase the uncertainties for each point in the transmission spectrum. As we are trying to find the relative change in $\rho$ with transmission spectroscopy (hence why we condition on fixed values of $T_\text{C}$, $a/R_\star$ and $b$), we tried to remove a common signal in all the light curves.

This common signal is evident in Fig.~\ref{fig:all_lightcurves}, where many of the residuals appear to have the same shape. This signal is also very similar to the GP model fitted to the white light curve (red dashed line, Fig.~\ref{fig:GP_white_lcv}). Taking this signal into account with the GP (or any other method) will increase the uncertainties in the calculated values of $\rho$, but this signal should not affect the relative values for $\rho$. We therefore divided through each spectral light curve by the GP systematics model fitted to the white light curve prior to fitting each light curve with the methods described above. This will remove the common signal and allow the noise models to produce more independent data points. Of course the underlying physical signal may not be identical for all the light curves, but the noise models described above can also take into account any excess signal added or not removed by the procedure in the same way that they account for `normal' systematics. In theory we could model this in more principled ways, e.g. by modelling all light curves simultaneously with a common signal plus independent signals, or perhaps by using the white light curve instrument model as an input for the systematics model, but we choose not to pursue them here and follow this simple procedure. Each fit for the spectral light curves as described in Sects.~\ref{sect:whiteanalysis}, \ref{sect:GPanalysis} and \ref{sect:waveletanalysis} was repeated by first dividing the light curve by the GP noise model for the white light curve. This procedure appeared to remove much of the common signal, and allowed for a more precise determination of the transmission spectrum, as described in the following section.

\section{Results and Conclusions}
\label{sect:results}

\subsection{White light curve analysis}
\label{sect:results_wlc}

The results from the white light curve fit are given in Tab.~\ref{tab:wlc_results} for three of the noise models used: white noise plus $\beta$ rescale, wavelets, and the Gaussian process after marginalising over all the transit parameters and covariance hyperparameters. The derived parameters are consistent with those reported in \citet{Hellier_2010} and \citet{Dragomir_2011}. In general, the three noise models gave consistent results. The simple white noise model produced the smallest uncertainties (which we do not reproduce here), whereas the white noise plus $\beta$ model gave the largest uncertainties. The wavelet and GP models both gave something in between. This is expected given that they both try to account for the form of the systematics model when inferring transit parameters, rather than just scaling the uncertainties to account for it. The fact that the wavelet and GP methods are not only consistent, but produce similar uncertainties is a strong validation of both techniques. We briefly discuss the relative merits of both methods in Sect.~\ref{sect:discussion}, and for the remainder of this paper we adopt the GP results, given that we cannot verify the $1/f$ nature of the noise, and they provide more conservative uncertainties. This maybe indicates that we are taking into account a larger range of possible systematic signals. The white noise values fitted for the GP and wavelet methods are $\sim$3.8 and 2.9 times the theoretical photon noise. The difference is perhaps due to the wavelet method absorbing some of the white noise into the systematic component (see e.g. Figs. 1 and 4 of \citealt{Carter_2009}, where the $1/f$ noise appears to contain a white component). Further sources of noise are accounted for by the time-correlated component of these models, given by the max covariance and red noise parameters quoted in Tab.~\ref{tab:wlc_results}.

Using the distributions from the MCMC chains, we calculate further system parameters of WASP-29b. These are given in Tab.~\ref{tab:wlc_results_derived}. Where the distributions were not available, we generated normal distributed values from the values given in the literature to propagate the uncertainties in the calculations properly. Again these results are consistent with the values reported in \citet{Hellier_2010} and \citet{Dragomir_2011} within the uncertainties, further confirming WASP-29b's status as a Saturn-like exoplanet.

\begin{table*}
\caption{Parameters from the MCMC fits of the white light curves, given for three different noise models: white noise plus $\beta$, wavelets, and the fully marginalised Mat\'ern 3/2 Gaussian process.}
\label{tab:wlc_results}
\begin{tabular}{llll}
\hline
\noalign{\smallskip}
~ & \multicolumn{3}{c}{Noise model}\\
Parameter & White Noise $+\beta$ & Wavelet & GP\\
\hline
\noalign{\smallskip}
Central transit time, $T_\text{C}$ (HJD$_{UTC}$) & $2455853.72515^{+0.00028}_{-0.00028}$ & $2455853.72469^{+0.00015}_{-0.00015}$ & $2455853.72442^{+0.00016}_{-0.00017}$ \\\noalign{\smallskip}
Period, $P$ (days) & $3.922727$ (fixed) & $3.922727$ (fixed) & $3.922727$ (fixed) \\\noalign{\smallskip}
System scale, $a/R_\star$ & $11.95^{+0.34}_{-0.67}$ & $12.36^{+0.12}_{-0.21}$ & $12.36^{+0.13}_{-0.22}$ \\\noalign{\smallskip}
Planet-star radius ratio, $R_\text{p}/R_\star$ & $0.0998^{+0.0026}_{-0.0017}$ & $0.0984^{+0.0010}_{-0.0009}$ & $0.0982^{+0.0015}_{-0.0015}$ \\\noalign{\smallskip}
Impact parameter, $b$ & $0.25^{+0.16}_{-0.16}$ & $0.14^{+0.10}_{-0.08}$ & $0.14^{+0.11}_{-0.09}$ \\\noalign{\smallskip}
Linear limb darkening parameter, $c_1$ & $0.721^{+0.076}_{-0.091}$ & $0.745^{+0.039}_{-0.047}$ & $0.698^{+0.059}_{-0.089}$ \\\noalign{\smallskip}
Quad limb darkening parameter, $c_2$ & $0.154^{+0.135}_{-0.094}$ & $0.090^{+0.095}_{-0.051}$ & $0.142^{+0.142}_{-0.084}$ \\\noalign{\smallskip}
Out-of-transit flux, $f_\text{oot}$ & $0.99969^{+0.00009}_{-0.00009}$ & $0.99963^{+0.00008}_{-0.00008}$ & $0.99965^{+0.00031}_{-0.00033}$ \\\noalign{\smallskip}
Time gradient, $T_\text{grad}$ & $-0.00014^{+0.00005}_{-0.00005}$ & $-0.00008^{+0.00004}_{-0.00003}$ & $-0.00008$ (fixed) \\\noalign{\smallskip}
GP max covariance, $\xi$ & $-$ & $-$ & $0.00058^{+0.00033}_{-0.00018}$ \\\noalign{\smallskip}
GP inverse length scale, $\eta$ & $-$ & $-$ & $25.7^{+14.9}_{-9.5}$ \\\noalign{\smallskip}
White noise, $\sigma_\text{w}$ & $0.00119$ (fixed) & $0.000324^{+0.000015}_{-0.000014}$ & $0.000426^{+0.000017}_{-0.000016}$ \\\noalign{\smallskip}
Red noise, $\sigma_\text{r}$ & $-$ & $0.00251^{+0.00033}_{-0.00030}$ & $-$ \\\noalign{\smallskip}\noalign{\smallskip}
\hline
\end{tabular}
\end{table*}

\begin{table}
\caption{WASP-29 parameters derived from the MCMC posterior distribution of the GP fits.}
\label{tab:wlc_results_derived}
\begin{tabular}{lll}
\hline
Parameter & Value & Unit \\
\hline
\noalign{\smallskip}
Transit epoch, $T_0$ & $2455830.18811^{+0.00016}_{-0.00016}$ & HJD$_{UTC}$ \\\noalign{\smallskip}
Period, $P$ & $3.9227186^{+0.0000068}_{-0.0000068}$ & days \\\noalign{\smallskip}
Transit duration, $T_{14}$ & $0.11036^{+0.00071}_{-0.00063}$ & days \\\noalign{\smallskip}
Inclination, $i$ & $89.17^{+0.50}_{-0.56}$ & deg \\\noalign{\smallskip}
Semi-major axis, $a$ & $0.04565^{+0.00060}_{-0.00062}$ & AU \\\noalign{\smallskip}
Stellar radius$^\alpha$, $R_\star$ & $0.808\pm0.044$ & $R_\odot$ \\\noalign{\smallskip}
Planet mass$^\alpha$, $M_\text{p}$ & $0.244\pm0.020$ & $M_J$ \\\noalign{\smallskip}
Planet radius, $R_\text{p}$ & $0.776^{+0.043}_{-0.043}$ & $R_J$ \\\noalign{\smallskip}
Planet density, $\rho_\text{p}$ & $0.53^{+0.11}_{-0.09}$ & $\rho_J$ \\\noalign{\smallskip}
Surface gravity, $\log g_\text{p}$ & $3.00^{+0.06}_{-0.06}$ & [cgs] \\\noalign{\smallskip}
Equilibrium temp, $T_\text{p}$ & $970^{+32}_{-31}$ & K \\\noalign{\smallskip}\hline
\noalign{\smallskip}
\multicolumn{3}{l}{{\footnotesize\,$^\alpha$\,Adopted from \citet{Hellier_2010}.}} \\
\noalign{\smallskip}
\end{tabular}
\end{table}

\subsection{Transit Ephemeris}

A new ephemeris was calculated for WASP-29 using the transit time derived for the white Gemini transit light curve, plus the ephemeris reported in \citet{Hellier_2010} and the transit time from \citet{Dragomir_2011}. These were converted to HJD$_{UTC}$ format, and a straight line of the form
\[
T_\text{C}(E) = T_\text{C}(0) + PE
\]
was fitted to the three transit times. The zero-point epoch was set equal to as near to the centre of mass of the three points as possible, weigthed as $1/\sigma_{T_\text{C}}^2$. This was to minimise the covariance between the transit epoch and the period, which was verified after the fit. The chosen $E=0$ transit was 6 periods prior to the Gemini transit, giving epochs of -130, -98 and 6 for the three transits. The new ephemeris is reported in Tab.~\ref{tab:wlc_results_derived}.

\subsection{Transmission spectrum}
\label{sect:trans_spectrum}

\begin{figure}
\includegraphics[width=85mm]{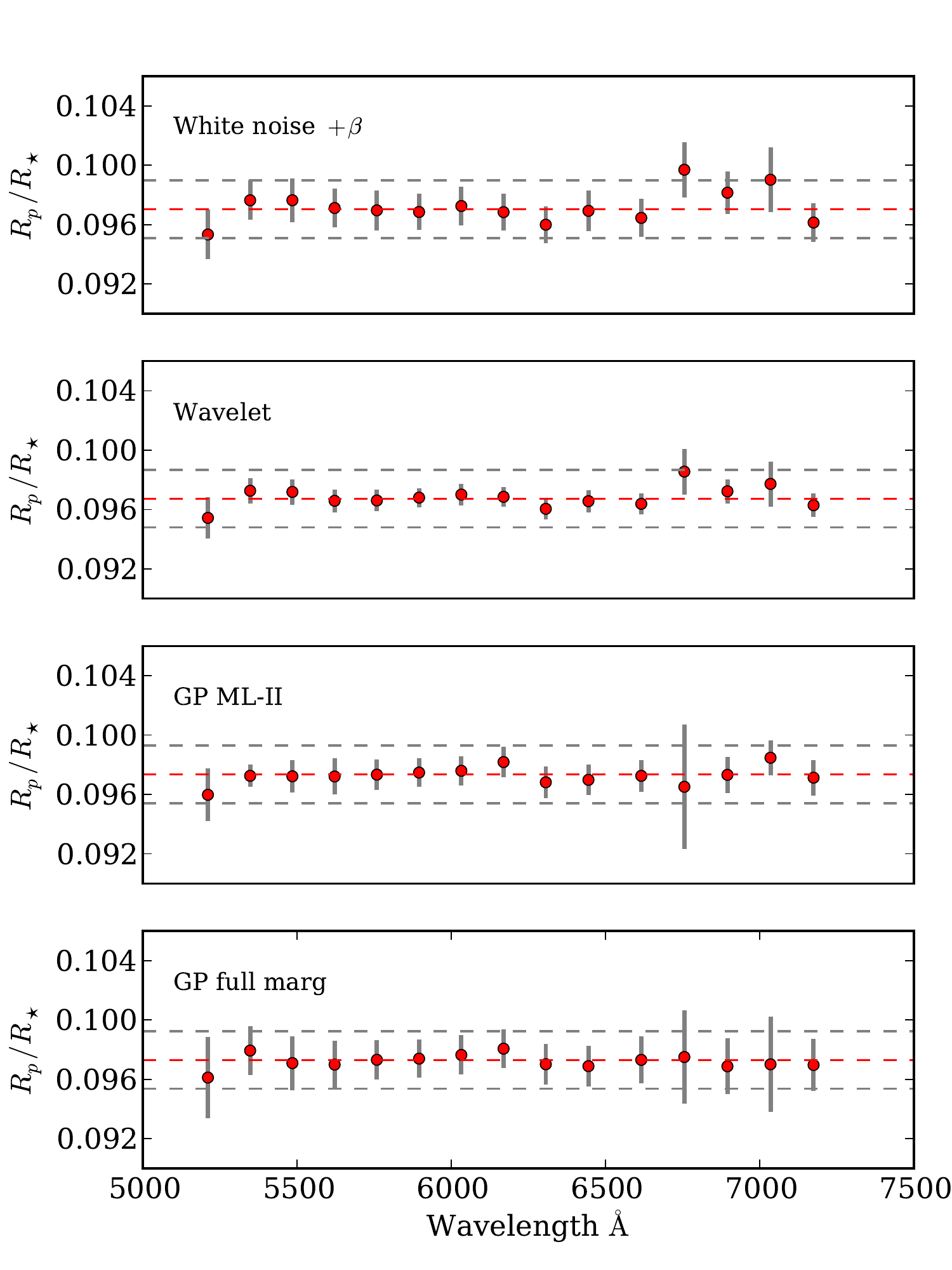}
\caption{Transmission spectra of WASP-29b produced by the various noise models prior to removal of the common mode systematic. The horizontal dashed lines represent the weighted average and plus and minus three scale heights, calculated to be $\sim$360\,km (one scale height of 360\,km corresponds to $\sim1.3\times10^{-4}$ in transit depth). In all cases the $\chi^2$ of a flat model is considerably lower than 1, indicating that the uncertainty in the relative planet-to-star radius ratio might be overestimated due to common mode systematics.}
\label{fig:transpec}
\end{figure}

The transmission spectra produced via the noise models discussed in Sect.~\ref{sect:analysis} are shown in Fig.~\ref{fig:transpec}. These are prior to removal of the common mode systematic. 
The horizontal dashed lines represent the weighted average and plus and minus three scale heights, calculated to be $\sim$360\,km (one scale height of 360\,km corresponds to $\sim1.3\times10^{-4}$ in transit depth). In general the transmission spectra are all in broad agreement, and are remarkably flat, showing a featureless spectrum at the few parts in $10^{-4}$ level. Uncertainties were smallest for the simple white noise model (not shown) and largest for the full GP marginalisation. However, the dispersion of the points (i.e. the scatter around the average) was smallest for the full GP marginalisation. This implies that the GP model is doing a particularly good job at determining the correct value for $\rho$, but for some reason overestimates the uncertainty (it is highly unlikely a draw from random noise would lead to such small dispersion). In fact, all of the noise models presented give reduced $\chi^2$ significantly smaller than 1. We propose that this is due to a common mode systematic that the noise models take into account in a similar way for each wavelength channel, and led to the correction for this as discussed in Sect.~\ref{sect:commonmode}.

\begin{figure}
\includegraphics[width=85mm]{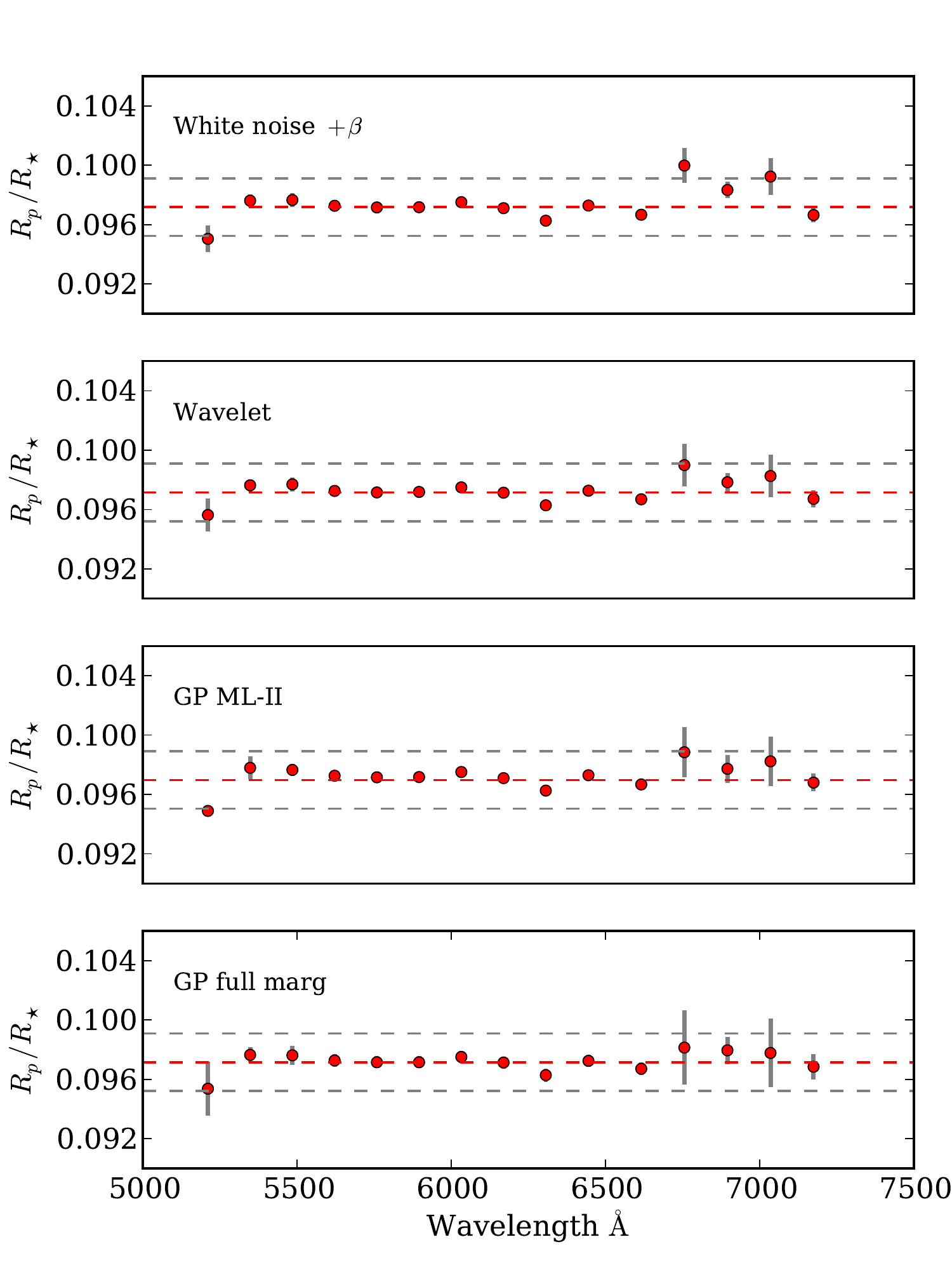}
\caption{Same as Fig.~\ref{fig:transpec} with the common mode systematic removed from all light curves prior to fitting.}
\label{fig:transpec_subsig}
\end{figure}

Fig.~\ref{fig:transpec_subsig} shows the transmission spectra produced after the common mode systematic correction. The spectra produced by all noise models were again consistent, with the full GP model giving uncertainties typically $30-40\%$ larger than the other models. We therefore choose to adopt these as our final uncertainties, for the reasons discussed in Sect.~\ref{sect:results_wlc}. These results are given in Tab.~\ref{tab:trans_spec_single}. The transmission spectrum is still consistent with a flat model, but we are able to place stronger constraints on the atmosphere of WASP-29b using the common mode correction. For the common mode corrected light curves, the white noise values fitted for the GP model ranges from 1.51 to 2.12 times the theoretical noise, and from 1.11 to 1.50 for the wavelet model. Similarly to the white light curve, further sources of noise are accounted for by the systematic component of these models. The lower ratios between the actual and theoretical white noise for the spectral light curves as compared to the white light curve perhaps indicate that atmospheric transmission corrections using the comparison star are best done in narrow wavelength ranges.

\begin{figure*}
\includegraphics[width=142mm]{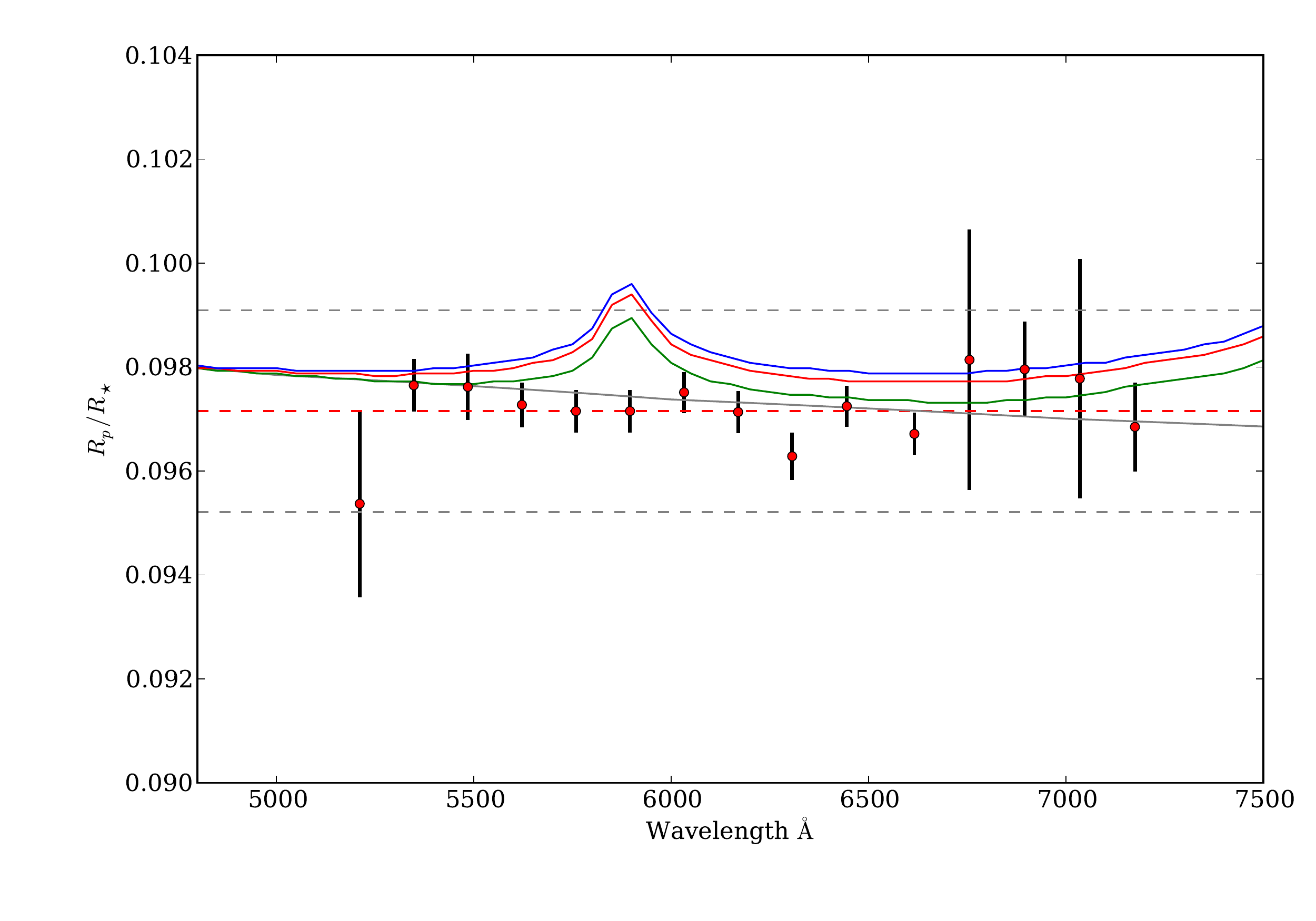}
\caption{Transmission spectrum of WASP-29b using the full Mat\'ern 3/2 GP noise model, and after removal of the common mode systematic. The horizontal dashed lines represent the weighted average and plus and minus three scale heights, calculated to be $\sim$360\,km (one scale height of 360\,km corresponds to $\sim1\times10^{-4}$ in transit depth).
The grey line shows a model containing a purely H$_2$ and He atmosphere. The green, red and blue lines are with 100 ppmv H$_2$O and 1, 5 and 10 ppmv of Na and K added, respectively.
These are not fitted to the data, but simply over plotted for reference. The models are plotted with a resolution of 50\AA; the points do not significantly change when plotted at the resolution of the spectrum.}
\label{fig:transpec_single}
\end{figure*}

Fig.~\ref{fig:transpec_single} shows the transmission spectrum produced with the full GP model, now with several model transmission spectra of WASP-29b overplotted. These forward models were produced using the {\sc nemesis} retrieval tool \citep{Irwin_2008},  a radiative transfer code originally developed to investigate the atmospheres of Solar system planets, and recently adapted for exoplanet transmission spectra \citep[][Barstow et al. {\it submitted}]{Lee_2012}. The grey line shows a model containing a purely H$_2$ and He atmosphere. The green, red and blue lines are with 100 ppmv H$_2$O and 1, 5 and 10 ppmv of Na and K added, respectively. Gas absorption line data is from \citet{Rothman_2010} and \citet{Kupka_2000} for the H$_2$O and the alkali metals, respectively. Given the precision of the transmission spectrum, we do not attempt a detailed retrieval here, rather the models are plotted for reference to show the scale of potential features.

Using our data we can only realistically rule out cloud-free atmospheres with significant amounts of Na, given the lack of a pressure broadened feature.
However, a Na rich atmosphere with thick clouds or Rayleigh scattering haze is not ruled out.
Of course, another explanation is that elemental Na is simply not present. This is likely for atmospheres of $\sim$1000\,K or cooler \citep[e.g.][]{Burrows_2000}, where atomic Na can be lost in compounds such as disodium monosulphide (Na$_2$S) or ansite (NaAlSi$_3$O$_8$), depending on the presence of other species in the atmosphere. Similar observations at shorter wavelengths could distinguish between a flat featureless spectrum and one dominated by a Rayleigh scattering haze, such as the prototypical HD 189733b \citep{Pont_2008,Sing_2012}, and higher resolution (at similar S/N) is required to rule out the presence of a Na core if there are clouds or haze present. For comparison the HD 189733b transmission spectrum varies by about 2 scale heights over this spectral range due to a Rayleigh scattering haze. Despite the lack of constraints we can place on the atmosphere, we can rule out WASP-29b having a similar atmosphere to the other prototypical hot Jupiter HD 209458b, given the lack of a strong pressure broadened Na feature. As stated, Na is likely to form compounds below $\sim$1000\,K. Given its prominence in HD 209458b's transmission spectrum, this could indicate another significant transition in classes of hot Jupiter atmospheres.

\begin{table}
\caption{Transmission spectrum of WASP-29b using the full GP marginalisation and after removal of the common mode systematic.}
\label{tab:trans_spec_single}
\begin{tabular}{ccc}
\hline
\noalign{\smallskip}
\noalign{\smallskip}
Wavelength & $\rho$ & $\Delta\rho$\\
\hline
5211.1\AA & 0.0961 & 0.0027\\
5348.0\AA & 0.0979 & 0.0016\\
5485.0\AA & 0.0971 & 0.0018\\
5621.9\AA & 0.0970 & 0.0016\\
5758.8\AA & 0.0973 & 0.0013\\
5895.8\AA & 0.0974 & 0.0013\\
6032.7\AA & 0.0976 & 0.0013\\
6169.6\AA & 0.0981 & 0.0013\\
6306.6\AA & 0.0970 & 0.0014\\
6444.9\AA & 0.0969 & 0.0014\\
6616.0\AA & 0.0973 & 0.0016\\
6755.7\AA & 0.0975 & 0.0032\\
6895.4\AA & 0.0969 & 0.0019\\
7035.0\AA & 0.0970 & 0.0032\\
7174.7\AA & 0.0970 & 0.0018\\
\noalign{\smallskip}
\hline
\end{tabular}
\end{table}

\section{Discussion}
\label{sect:discussion}

We have presented Gemini GMOS observations of the transmission spectrum of WASP-29b, using the technique of differential spectro-photometry. Using a single comparison star, we reached precision on the transit depth of $\sim1\times10^{-4}$ showing that GMOS can provide precision spectrophotometry at the level needed to probe the atmospheres of extrasolar planets.

Using the `white' light curve, we refined the system parameters and ephemeris for WASP-29, finding them to be consistent with previous studies \citep{Hellier_2010,Dragomir_2011}.
Despite picking WASP-29 as a test case for GMOS transmission spectroscopy, the precision attained allows us to rule out a Na rich, cloud/haze free atmosphere, given the lack of a pressure broadened Na feature. This indicates that Na is not present in the atmospheres of cooler `hot' Jupiters, or that clouds and/or hazes play an important role and mask the pressure broadened alkali metal signatures in WASP-29b's upper atmosphere. The former explanation is perhaps more likely, although a spectrum covering a larger wavelength region is required to confirm this, and it is of course possible that both are true. A higher resolution spectra (at similar S/N) is required to rule out the presence of a narrow Na core if clouds or hazes dominate. We note that this represents the first transmission spectrum of a hot-Saturn planet.

We have also presented a detailed analysis and comparison using various types of noise models to account for the GMOS systematics. Rather than search for correlations with observational parameters such as seeing, airmass, etc., we decided to focus on blind methods to account for the systematics, i.e. with no additional input parameters used to model the systematics. The methods tried include a white noise model, a simple rescaling of the photometric uncertainties, the wavelet method of \citet{Carter_2009}, and the Gaussian process model of \citet{Gibson_2012} applied to time correlated noise. The more sophisticated methods gave similar uncertainties, verifying their usefulness for analysis of time-correlated systematics.

In general, we restate some of the conclusions of \citet{Carter_2009}, that any model taking into account time-correlated noise is better than ignoring it, and that an analysis of the residuals using ACFs and PSDs are especially useful in guiding the choice of noise model required.
The wavelet and GP models give similar results and uncertainties for the white light curve, and within about $40\%$ for the uncertainties in the transmission spectrum. Despite GPs giving a slightly more conservative estimate of the uncertainties for the GMOS data, the wavelet method is perhaps preferred in general for the analysis of time-correlated noise when the PSD of the residuals follow a $1/f^\gamma$ distribution, given its much faster execution time (although we note that this is hard to verify for individual datasets, and care must be taken for low significance results). The GP method is more general, and can be applied to almost any noise model given a suitable choice of kernel (potentially even non-stationary noise), non-regularly spaced data, and can incorporate arbitrary numbers of input vectors into the stochastic function. This allows physical systematics models to be folded into the stochastic part of the GP, therefore allowing principled Bayesian inference of the instrument model and negating the need to specify the instrument model in closed form \citep{Gibson_2012,Gibson_2012b}. However, this added functionality comes at a significant runtime cost, and restricts the use of GPs up to datasets of $\sim$1000 points (at least using full marginalisation over the hyperparmeters with MCMC methods). Investigations into sparse GP models may allow their application to larger datasets \citep[e.g.][]{quinonero2005unifying}. We finally note that given the difficulty in dealing with systematic noise, the use of multiple, complimentary techniques is desirable where possible, although perhaps the only truly robust way to confirm results is to repeat measurements.

\section*{Acknowledgments}

Based on observations obtained at the Gemini Observatory, which is operated by the Association of Universities for Research in Astronomy (AURA) under a cooperative agreement with the NSF on behalf of the Gemini partnership: the National Science Foundation (United States), the Science and Technology Facilities Council (United Kingdom), the National Research Council (Canada), CONICYT (Chile), the Australian Research Council (Australia), CNPq (Brazil) and CONICET (Argentina). We are extremely grateful for the support provided by the Gemini staff. We also thank the referee for comments that improved the paper.
N. P. G and S. A. acknowledge support from STFC grant ST/G002266/2. J. K. B. acknowledges the support of the John Fell Oxford University Press (OUP) Research Fund for this research. L.N.F. was supported by a Glasstone Fellowship at the University of Oxford.

\bibliography{../MyBibliography} 

\begin{thebibliography}{32}
\expandafter\ifx\csname natexlab\endcsname\relax\def\natexlab#1{#1}\fi

\bibitem[{{Bean} {et~al.}(2011){Bean}, {D{\'e}sert}, {Kabath}, {Stalder},
  {Seager}, {Miller-Ricci Kempton}, {Berta}, {Homeier}, {Walsh}, \&
  {Seifahrt}}]{Bean_2011}
{Bean} J.~L., {D{\'e}sert} J.-M., {Kabath} P., {Stalder} B., {Seager} S.,
  {Miller-Ricci Kempton} E., {Berta} Z.~K., {Homeier} D., {Walsh} S.,
  {Seifahrt} A., 2011, \apj, 743, 92

\bibitem[{{Bean} {et~al.}(2010){Bean}, {Miller-Ricci Kempton}, \&
  {Homeier}}]{Bean_2010}
{Bean} J.~L., {Miller-Ricci Kempton} E., {Homeier} D., 2010, \nat, 468, 669

\bibitem[{{Berta} {et~al.}(2012){Berta}, {Charbonneau}, {D{\'e}sert},
  {Miller-Ricci Kempton}, {McCullough}, {Burke}, {Fortney}, {Irwin}, {Nutzman},
  \& {Homeier}}]{Berta_2012}
{Berta} Z.~K., {Charbonneau} D., {D{\'e}sert} J.-M., {Miller-Ricci Kempton} E.,
  {McCullough} P.~R., {Burke} C.~J., {Fortney} J.~J., {Irwin} J., {Nutzman} P.,
  {Homeier} D., 2012, \apj, 747, 35

\bibitem[{{Brown}(2001)}]{Brown_2001}
{Brown} T.~M., 2001, \apj, 553, 1006

\bibitem[{{Burrows} {et~al.}(2000){Burrows}, {Marley}, \&
  {Sharp}}]{Burrows_2000}
{Burrows} A., {Marley} M.~S., {Sharp} C.~M., 2000, \apj, 531, 438

\bibitem[{{Carter} \& {Winn}(2009)}]{Carter_2009}
{Carter} J.~A., {Winn} J.~N., 2009, \apj, 704, 51

\bibitem[{{Charbonneau} {et~al.}(2002){Charbonneau}, {Brown}, {Noyes}, \&
  {Gilliland}}]{Charbonneau_2002}
{Charbonneau} D., {Brown} T.~M., {Noyes} R.~W., {Gilliland} R.~L., 2002, \apj,
  568, 377

\bibitem[{{Dragomir} {et~al.}(2011){Dragomir}, {Kane}, {Pilyavsky},
  {Mahadevan}, {Ciardi}, {Gazak}, {Gelino}, {Payne}, {Rabus}, {Ramirez}, {von
  Braun}, {Wright}, \& {Wyatt}}]{Dragomir_2011}
{Dragomir} D., {Kane} S.~R., {Pilyavsky} G., {Mahadevan} S., {Ciardi} D.~R.,
  {Gazak} J.~Z., {Gelino} D.~M., {Payne} A., {Rabus} M., {Ramirez} S.~V., {von
  Braun} K., {Wright} J.~T., {Wyatt} P., 2011, \aj, 142, 115

\bibitem[{{Gelman} \& {Rubin}(1992)}]{GelmanRubin_1992}
{Gelman} A., {Rubin} D.~B., 1992, Stat. Sci., 7, 457

\bibitem[{{Gibson} {et~al.}(2012{\natexlab{a}}){Gibson}, {Aigrain}, {Pont},
  {Sing}, {D{\'e}sert}, {Evans}, {Henry}, {Husnoo}, \&
  {Knutson}}]{Gibson_2012b}
{Gibson} N.~P., {Aigrain} S., {Pont} F., {Sing} D.~K., {D{\'e}sert} J.-M.,
  {Evans} T.~M., {Henry} G., {Husnoo} N., {Knutson} H., 2012{\natexlab{a}},
  \mnras, 422, 753

\bibitem[{{Gibson} {et~al.}(2012{\natexlab{b}}){Gibson}, {Aigrain}, {Roberts},
  {Evans}, {Osborne}, \& {Pont}}]{Gibson_2012}
{Gibson} N.~P., {Aigrain} S., {Roberts} S., {Evans} T.~M., {Osborne} M., {Pont}
  F., 2012{\natexlab{b}}, \mnras, 419, 2683

\bibitem[{{Gibson} {et~al.}(2009){Gibson}, {Pollacco}, {Simpson}, {Barros},
  {Joshi}, {Todd}, {Keenan}, {Skillen}, {Benn}, {Christian}, {Hrudkov{\'a}}, \&
  {Steele}}]{Gibson_2009}
{Gibson} N.~P., {Pollacco} D., {Simpson} E.~K., {Barros} S., {Joshi} Y.~C.,
  {Todd} I., {Keenan} F.~P., {Skillen} I., {Benn} C., {Christian} D.,
  {Hrudkov{\'a}} M., {Steele} I.~A., 2009, \apj, 700, 1078

\bibitem[{{Gibson} {et~al.}(2008){Gibson}, {Pollacco}, {Simpson}, {Joshi},
  {Todd}, {Benn}, {Christian}, {Hrudkov{\'a}}, {Keenan}, {Meaburn}, {Skillen},
  \& {Steele}}]{Gibson_2008}
{Gibson} N.~P., {Pollacco} D., {Simpson} E.~K., {Joshi} Y.~C., {Todd} I.,
  {Benn} C., {Christian} D., {Hrudkov{\'a}} M., {Keenan} F.~P., {Meaburn} J.,
  {Skillen} I., {Steele} I.~A., 2008, \aap, 492, 603

\bibitem[{{Gibson} {et~al.}(2011){Gibson}, {Pont}, \& {Aigrain}}]{Gibson_2011}
{Gibson} N.~P., {Pont} F., {Aigrain} S., 2011, \mnras, 411, 2199

\bibitem[{{Hellier} {et~al.}(2010){Hellier}, {Anderson}, {Collier Cameron},
  {Gillon}, {Lendl}, {Maxted}, {Queloz}, {Smalley}, {Triaud}, {West}, {Brown},
  {Enoch}, {Lister}, {Pepe}, {Pollacco}, {S{\'e}gransan}, \&
  {Udry}}]{Hellier_2010}
{Hellier} C., {Anderson} D.~R., {Collier Cameron} A., {Gillon} M., {Lendl} M.,
  {Maxted} P.~F.~L., {Queloz} D., {Smalley} B., {Triaud} A.~H.~M.~J., {West}
  R.~G., {Brown} D.~J.~A., {Enoch} B., {Lister} T.~A., {Pepe} F., {Pollacco}
  D., {S{\'e}gransan} D., {Udry} S., 2010, \apjl, 723, L60

\bibitem[{{Huitson} {et~al.}(2012){Huitson}, {Sing}, {Vidal-Madjar},
  {Ballester}, {Lecavelier des Etangs}, {D{\'e}sert}, \& {Pont}}]{Huitson_2012}
{Huitson} C.~M., {Sing} D.~K., {Vidal-Madjar} A., {Ballester} G.~E.,
  {Lecavelier des Etangs} A., {D{\'e}sert} J.-M., {Pont} F., 2012, \mnras, 422,
  2477

\bibitem[{{Irwin} {et~al.}(2008){Irwin}, {Teanby}, {de Kok}, {Fletcher},
  {Howett}, {Tsang}, {Wilson}, {Calcutt}, {Nixon}, \& {Parrish}}]{Irwin_2008}
{Irwin} P.~G.~J., {Teanby} N.~A., {de Kok} R., {Fletcher} L.~N., {Howett}
  C.~J.~A., {Tsang} C.~C.~C., {Wilson} C.~F., {Calcutt} S.~B., {Nixon} C.~A.,
  {Parrish} P.~D., 2008, J. Quant. Spectrosc. Radiative Transfer, 109, 1136

\bibitem[{{Kupka} {et~al.}(2000){Kupka}, {Ryabchikova}, {Piskunov}, {Stempels},
  \& {Weiss}}]{Kupka_2000}
{Kupka} F.~G., {Ryabchikova} T.~A., {Piskunov} N.~E., {Stempels} H.~C., {Weiss}
  W.~W., 2000, Baltic Astronomy, 9, 590

\bibitem[{{Lee} {et~al.}(2012){Lee}, {Fletcher}, \& {Irwin}}]{Lee_2012}
{Lee} J.-M., {Fletcher} L.~N., {Irwin} P.~G.~J., 2012, \mnras, 420, 170

\bibitem[{{Mandel} \& {Agol}(2002)}]{mandel_agol_2002}
{Mandel} K., {Agol} E., 2002, \apjl, 580, L171

\bibitem[{{Pont} {et~al.}(2008){Pont}, {Knutson}, {Gilliland}, {Moutou}, \&
  {Charbonneau}}]{Pont_2008}
{Pont} F., {Knutson} H., {Gilliland} R.~L., {Moutou} C., {Charbonneau} D.,
  2008, \mnras, 385, 109

\bibitem[{{Pont} {et~al.}(2006){Pont}, {Zucker}, \& {Queloz}}]{Pont_2006}
{Pont} F., {Zucker} S., {Queloz} D., 2006, \mnras, 373, 231

\bibitem[{Qui{\~n}onero-Candela \& Rasmussen(2005)}]{quinonero2005unifying}
Qui{\~n}onero-Candela J., Rasmussen C., 2005, The Journal of Machine Learning
  Research, 6, 1939

\bibitem[{{Rasmussen} \& {Williams}(2006)}]{Rasmussen_Williams}
{Rasmussen} C.~E., {Williams} K.~I., 2006, {Gaussian Processes for Machine
  Learning}. {The MIT Press}

\bibitem[{{Rothman} {et~al.}(2010){Rothman}, {Gordon}, {Barber}, {Dothe},
  {Gamache}, {Goldman}, {Perevalov}, {Tashkun}, \& {Tennyson}}]{Rothman_2010}
{Rothman} L.~S., {Gordon} I.~E., {Barber} R.~J., {Dothe} H., {Gamache} R.~R.,
  {Goldman} A., {Perevalov} V.~I., {Tashkun} S.~A., {Tennyson} J., 2010,
  {Journal of Quantitative Spectroscopy and Radiative Transfer}, 111, 2139

\bibitem[{{Seager} \& {Sasselov}(2000)}]{Seager_2000}
{Seager} S., {Sasselov} D.~D., 2000, \apj, 537, 916

\bibitem[{{Sing} {et~al.}(2012){Sing}, {Huitson}, {Lopez-Morales}, {Pont},
  {D{\'e}sert}, {Ehrenreich}, {Wilson}, {Ballester}, {Fortney}, {Lecavelier des
  Etangs}, \& {Vidal-Madjar}}]{Sing_2012}
{Sing} D.~K., {Huitson} C.~M., {Lopez-Morales} M., {Pont} F., {D{\'e}sert}
  J.-M., {Ehrenreich} D., {Wilson} P.~A., {Ballester} G.~E., {Fortney} J.~J.,
  {Lecavelier des Etangs} A., {Vidal-Madjar} A., 2012, ArXiv e-prints

\bibitem[{{Sing} {et~al.}(2011){Sing}, {Pont}, {Aigrain}, {Charbonneau},
  {D{\'e}sert}, {Gibson}, {Gilliland}, {Hayek}, {Henry}, {Knutson}, {Lecavelier
  Des Etangs}, {Mazeh}, \& {Shporer}}]{Sing_2011}
{Sing} D.~K., {Pont} F., {Aigrain} S., {Charbonneau} D., {D{\'e}sert} J.-M.,
  {Gibson} N., {Gilliland} R., {Hayek} W., {Henry} G., {Knutson} H.,
  {Lecavelier Des Etangs} A., {Mazeh} T., {Shporer} A., 2011, \mnras, 416, 1443

\bibitem[{{Sing} {et~al.}(2008){Sing}, {Vidal-Madjar}, {D{\'e}sert},
  {Lecavelier des Etangs}, \& {Ballester}}]{Sing_2008}
{Sing} D.~K., {Vidal-Madjar} A., {D{\'e}sert} J.-M., {Lecavelier des Etangs}
  A., {Ballester} G., 2008, \apj, 686, 658

\bibitem[{{Swain} {et~al.}(2008){Swain}, {Vasisht}, \& {Tinetti}}]{Swain_2008}
{Swain} M.~R., {Vasisht} G., {Tinetti} G., 2008, \nat, 452, 329

\bibitem[{{Waldmann}(2012)}]{Waldmann_2012}
{Waldmann} I.~P., 2012, \apj, 747, 12

\bibitem[{{Winn} {et~al.}(2008){Winn}, {Holman}, {Torres}, {McCullough},
  {Johns-Krull}, {Latham}, {Shporer}, {Mazeh}, {Garcia-Melendo}, {Foote},
  {Esquerdo}, \& {Everett}}]{Winn_2008}
{Winn} J.~N., {Holman} M.~J., {Torres} G., {McCullough} P., {Johns-Krull} C.,
  {Latham} D.~W., {Shporer} A., {Mazeh} T., {Garcia-Melendo} E., {Foote} C.,
  {Esquerdo} G., {Everett} M., 2008, \apj, 683, 1076

\end{thebibliography}
\bibliographystyle{mn2e_astronat} 

\appendix
\section{GP trials on simulated light curves}
\label{appendix}

In order to choose the best GP kernel for time-correlated noise, we ran a series of tests on simulated light curves, with injected `systematic' noise. This appendix briefly describes our results.

In total we simulated 2\,400 light curves with 250 data points. For each light curve we set the transit parameters as follows: $P$ = 4.0\,days, $a/R_\star$ = 12.0, $\rho$ = 0.1, $b$ = 0.25, $c_1$ = 0.2, and $c_2$ = 0.2. A white and red noise term were then picked from a uniform distribution between 0.0001 to 0.0006. The injected systematic signal was simulated in a variety of ways. First, we created `function noise', where we summed 100 exponential, Gaussian and sinusoidal functions with random parameters (within sensible limits), and rescaled so that the mean and standard deviation were equal to unity and the red noise term, respectively. Second, we created $1/f$ noise in a similar way to \citet{Carter_2009}. We created a signal in the Fourier domain corresponding to a PSD of $1/f$, by setting a random amplitude within a $1/f^{0.5}$ envelope (the PSD is the square of the Fourier transform magintude), with a corresponding random phase. The inverse Fourier transform then produced the systematic signal. The signal was then scaled to have mean and standard deviation in the same way as before. Finally, we created some signals using a combination of these methods, by simply adding the two signals each with individual red noise terms chosen (and rescaling to a mean of 1). The model light curve was multiplied by the systematic signal, and the white noise was finally added. Out of the 2\,400 light curves, 800 were created using `function noise', 800 with $1/f$ noise, and 800 with the combined noise model. The light curves were inspected to ensure they appeared realistic.

Each light curve was fitted using a white noise model, as described in Sect.~\ref{sect:whiteanalysis}, and the Gaussian process model described in Sect.~\ref{sect:GPanalysis}, this time using a range of kernels, including the squared exponential (SE), rational quadratic (RQ) and Mat\'ern 3/2 (MAT). The MAT kernel is already defined in Eq.~\ref{eq:kernel}. The SE kernel is:
\[
k({t}_n, {t}_m | \btheta) = \xi^2 \exp \left( -\eta\Delta t^2\right) + \delta_{nm}\sigma_\text{w}^2.
\]
This kernel has the same parameters as the MAT kernel, only the shape changes. The MAT kernel is more sharply peaked at $\Delta t$ = 0, resulting in a rougher function, whereas the SE is an infinitely differentiable, smooth function of the input.
The RQ kernel is:
\[
k({t}_n, {t}_m | \btheta) = \xi^2 \left( 1+ \frac{\Delta t^2}{2\alpha l^2}   \right)^{-\alpha} + \delta_{nm}\sigma_\text{w}^2,
\]
where $\alpha$ is a shape parameter and $l$ is a length scale parameter. This kernel is a smooth function, and is a scale mixture of SE kernels with different characteristic length scales. The limit as $\alpha\rightarrow\infty$ is the SE kernel. For a detailed discussion of GP kernels see \citet{Rasmussen_Williams}.

\begin{figure*}
\includegraphics[width=142mm]{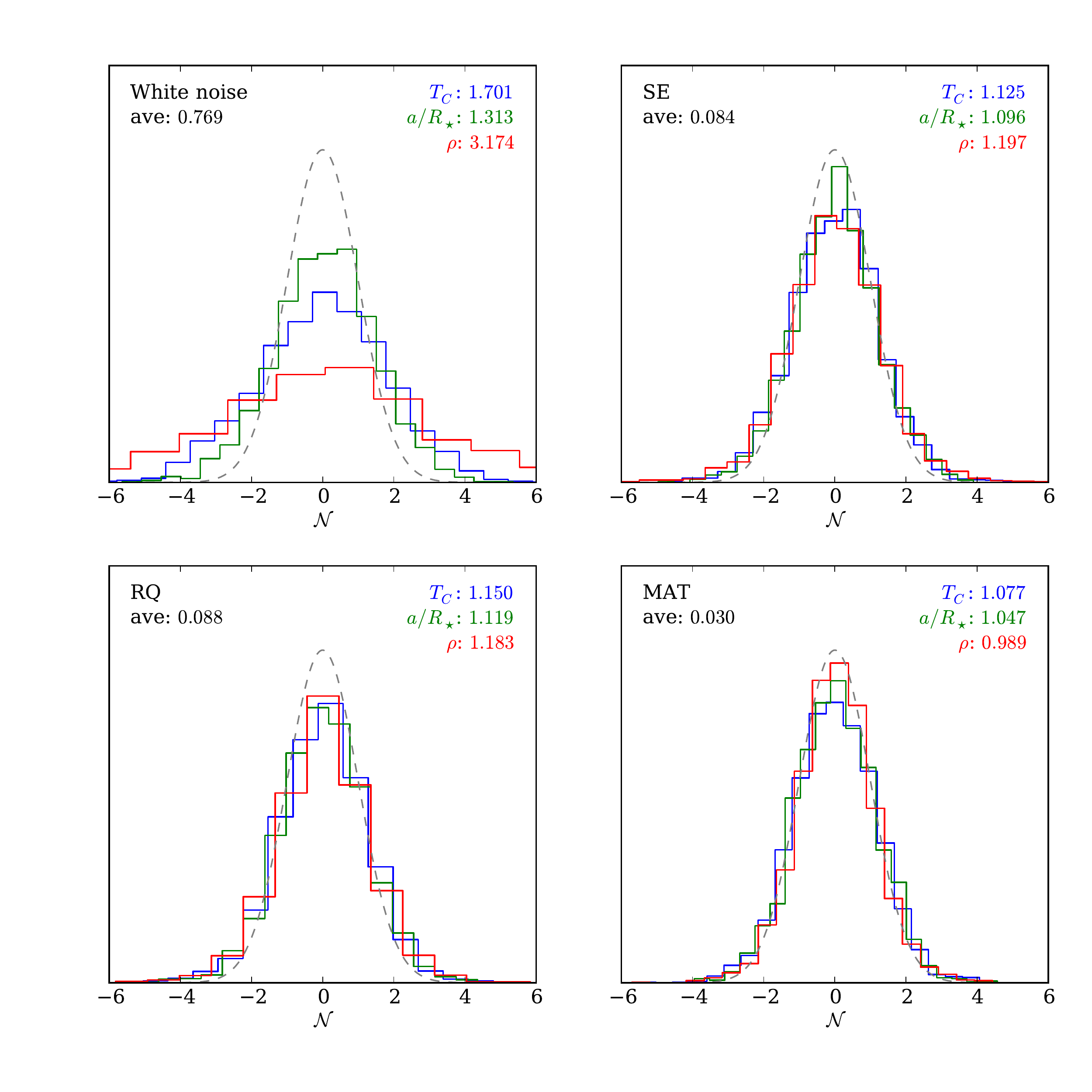}
\caption{Histograms of the `number-of-sigma' statistic for each of the noise models used. The plots are colour coded and show the distributions for the central transit time, system scale, and the planet-to-star radius ratio. The standard deviation of each is given, along with the average (see text). The dashed line shows a Gaussian with a mean of zero and standard deviation of 1, i.e. what an ideal noise model would produce.}
\label{fig:appendix_plot}
\end{figure*}

We ran a single MCMC chain of length 5000 for each fit, and discarded the first 1000 points. We checked for convergence visually for a subset of the light curves. We fitted for $T_\text{C}$, $\rho$, $a/R_\star$ and $f_\text{oot}$, and the kernel hyperparameters. $b$, $c_1$, $c_2$ and $T_\text{grad}$ were held fixed. This was a compromise to maintain degeneracies in the fit, but also to allow for shorter MCMC chains and therefore substantial numbers of trial light curves. We followed the approach of \citet{Carter_2009} to analyse the results. We calculated the `number-of-sigma' statistic for each parameter,
\[
\mathcal{N} = (\hat p- p)/\sigma_p,
\]
where $\hat p$ is the parameter estimate from the MCMC chain, $\sigma_p$ is the uncertainty, and $p$ is the true parameter value. This statistic should be distributed with a mean and variance of 0 and 1, respectively, if the parameter uncertainties from the model fits are Gaussian with the derived uncertainty. The results are plotted in Fig.~\ref{fig:appendix_plot} for the white noise model and the three GP models. The standard deviation for the fitted parameters are given in each plot, along with their average distance from 1.0 (calculated as the standard deviation with fixed mean of 1.0).

These results show that the MAT kernel outperforms the others for analysis of time-correlated noise, therefore we selected it for the GMOS analysis. However, we note that this is only valid for the specific noise models we tried, and varies significantly with varying noise parameters. 
We also note that these results are likely to depend in a complex way on the parameters of the light curve and the number of data points, as well as the injected noise properties. These tests were designed as a simple way of choosing the best kernel for time-correlated noise like we see in the GMOS light curves, and are not intended to be complete. Indeed, the right kernel to use is probably best selected on a case-by-case basis. Perhaps most importantly, we note that all three kernels invariably gave results significantly better than the simple white noise model. This demonstrates that any reasonably chosen kernel performs better than a simple white noise analysis when time-correlated noise is present.

\label{lastpage}

\end{document}